\documentclass[twocolumn,twocolappendix,times,tighten]{aastex631}

\usepackage{amsmath}  
\usepackage{bm}  
\usepackage{enumitem}  

\usepackage{color}

\newcommand*{\mt}{\mathrm}
\newcommand*{\unit}[1]{\;\mt{#1}}  
\newcommand*{\abt}{\mathord{\sim}} 
\newcommand*{\ptl}{\partial}
\newcommand*{\dtl}{\mathrm{d}}

\renewcommand{\vec}[1]{\boldsymbol{#1}}  

\newcommand*{\bp}{\beta_{\mathrm{p}}}
\newcommand*{\bpo}{\beta_{\mathrm{p0}}}
\newcommand*{\kB}{k_{\mathrm{B}}}
\newcommand*{\prll}{\parallel}
\newcommand*{\vA}{v_{\mathrm{A}}}  
\newcommand*{\vAo}{v_{\mathrm{A0}}}  

\newcommand*{\me}{m_{\mathrm{e}}}  
\newcommand*{\mi}{m_{\mathrm{i}}}
\newcommand*{\mime}{m_{\mathrm{i}}/m_{\mathrm{e}}}  

\newcommand*{\Te}{T_{\mathrm{e}}}  
\newcommand*{\Ti}{T_{\mathrm{i}}}

\newcommand*{\ompi}{\omega_{\mathrm{pi}}}
\newcommand*{\ompeo}{\omega_{\mathrm{pe0}}}
\newcommand*{\ompio}{\omega_{\mathrm{pi0}}}
\newcommand*{\Omce}{\Omega_{\mathrm{e}}}  
\newcommand*{\Omci}{\Omega_{\mathrm{i}}}
\newcommand*{\Omcio}{\Omega_{\mathrm{i0}}}

\newcommand*{\rLi}{\rho_{\mathrm{i}}}
\newcommand*{\rLio}{\rho_{\mathrm{i0}}}

\shorttitle{ICM Ion Cyclotron Scattering}  
\shortauthors{Tran et al.}  
\received{2022 September 26}
\revised{2023 February 8}
\accepted{2023 February 20}
\submitjournal{ApJ}

\begin{document}

\title{Electron Re-acceleration via Ion Cyclotron Waves in the Intracluster Medium}

\correspondingauthor{Aaron Tran, Lorenzo Sironi}
\email{aaron.tran@columbia.edu, lsironi@astro.columbia.edu}

\author[0000-0003-3483-4890]{Aaron Tran}
\affiliation{Department of Astronomy, Columbia University,
550 W 120th St.~MC~5246, New York, NY 10027, USA}

\author[0000-0002-5951-0756]{Lorenzo Sironi}
\affiliation{Department of Astronomy, Columbia University,
550 W 120th St.~MC~5246, New York, NY 10027, USA}

\author[0000-0002-8820-8177]{Francisco Ley}
\affiliation{Department of Astronomy, University of Wisconsin-Madison,
475 N Charter St., Madison, WI 53706, USA}

\author[0000-0003-4821-713X]{Ellen G.~Zweibel}
\affiliation{Department of Astronomy, University of Wisconsin-Madison,
475 N Charter St., Madison, WI 53706, USA}
\affiliation{Department of Physics, University of Wisconsin-Madison,
1150 University Ave., Madison, WI 53706, USA}

\author[0000-0003-2928-6412]{Mario A.~Riquelme}
\affiliation{Departamento de F\'{i}sica,
Facultad de Ciencias F\'{i}sicas y Matem\'{a}ticas,
Universidad de Chile, Av.~Blanco Encalada 2008, Santiago, Chile}

\begin{abstract}
In galaxy clusters, the intracluster medium (ICM) is expected to host a
diffuse, long-lived, and invisible population of ``fossil'' cosmic-ray
electrons (CRe) with 1--100 MeV energies.
These CRe, if re-accelerated by 100x in energy, can contribute synchrotron
luminosity to cluster radio halos, relics, and phoenices.
Re-acceleration may be aided by CRe scattering upon the ion-Larmor-scale waves
that spawn when ICM is compressed, dilated, or sheared.
We study CRe scattering and energy gain due to ion cyclotron (IC) waves
generated by continuously-driven compression in 1D fully kinetic
particle-in-cell simulations.
We find that pitch-angle scattering of CRe by IC waves induces energy gain via
magnetic pumping.
In an optimal range of IC-resonant momenta, CRe may gain up to
$\abt 10$--$30\%$ of their initial energy in one compress/dilate cycle with
magnetic field amplification $\sim 3$--$6\times$, assuming adiabatic
decompression without further scattering and averaging over initial pitch
angle.
\end{abstract}

\keywords{Plasma astrophysics (1261), Intracluster medium (858),
Cosmic rays (329), Non-thermal radiation sources (1119)}

\section{Introduction}

Clusters of galaxies host hot, diffuse, X-ray emitting gas which we call the
intracluster medium (ICM).
Some clusters, especially disturbed and merging clusters, also host a rich
variety of diffuse MHz--GHz radio emission in their ICM: radio synchrotron
halos, bridges, relics, and phoenices powered by relativistic cosmic-ray
electrons (CRe) \citep{van-weeren2019}.
These CRe cool via synchrotron radiation and inverse-Compton scattering off
cosmic microwave background photons over Megayears to Gigayears, reaching
1--100 MeV energies.
Because radiative power losses decrease at lower electron energies, and Coulomb
collisions are weak in the ICM, MeV ``fossil'' CRe may persist in clusters for
$\gtrsim$ Gigayears \citep{ensslin1999,petrosian2001,pinzke2013}.

Fossil CRe energies are too low to emit detectable radio synchrotron emission.
But, a re-acceleration of $100\times$ in energy can make fossil CRe shine again
in radio synchrotron and permit them to contribute to the power budget of radio
emission in the ICM \citep{brunetti2001,van-weeren2019,brunetti2020}.
Many mechanisms can energize fossil CRe:
large-scale adiabatic compression from sub-sonic sloshing or shocks
\citep{ensslin2001,markevitch2005},
diffusive shock acceleration in cluster merger shocks
\citep{kang2012-relic,guo2014-accel,kang2016-model,van-weeren2017,ha2022},
and wave damping or reconnection within a turbulent scale-by-scale cascade
\citep{brunetti2007,brunetti2011,brunetti2016}.

We consider another possibility for re-accelerating fossil CRe, wherein
large-scale deformation---compression, dilation, or shear---drives small-scale
plasma waves that might scatter and energize CRe directly.
When the ICM deforms on timescales shorter than the
Coulomb collision time and
longer than the Larmor gyration time, the $\vec{B}$-perpendicular
temperature $T_\perp$ changes due to conservation
of particle magnetic moment ${p_\perp}^2/B$, and the $\vec{B}$-parallel
temperature $T_\prll$ changes due to conservation of particle
bounce invariant $\oint p_\prll \dtl s$ integrated along a field line
(assuming periodicity in parallel motion).
As $T_\perp$ and $T_\prll$ evolve independently, the plasma becomes temperature
and pressure anisotropic: $\Delta \equiv T_\perp/T_\prll - 1 \ne 0$.
Because the ICM's thermal pressure dominates over magnetic pressure, i.e., its
plasma beta $\bp = P_\mt{thermal}/P_\mt{magnetic} \gtrsim 1$, $\Delta \ne 0$
easily triggers the growth of various Larmor-scale plasma waves
\citep{kasper2002,bale2009,kunz2014,kunz2019}.
The strongest waves reside at proton Larmor scales; although they are triggered
by and regulated by proton anisotropy, they may also interact with fossil CRe,
which gyrate more slowly and have larger Larmor radii than typical ICM thermal
electrons.

We focus on CRe interaction with ion cyclotron (IC) waves driven
by thermal ICM proton (i.e., ion) anisotropy $\Delta > 0$, with the anisotropy
in turn driven by continuous compression.
IC waves interact with electrons via the gyro-resonance condition:
\begin{equation} \label{eq:res}
    \omega - k v_\prll = - |\Omce|/\gamma ,
\end{equation}
where $\omega$ is wave angular frequency, $k = 2\pi/\lambda$ is wavenumber,
$\lambda$ is wavelength, $v_\prll$ is electron velocity parallel to $\vec{B}$,
$\Omce = -eB/(\me c)$ is the signed, non-relativistic electron cyclotron
frequency, and $\gamma$ is the electron's Lorentz factor.
Eq.~\eqref{eq:res} specifies an ``anomalous'' resonance, wherein an electron
overtaking the wave ($|v_\prll| > |\omega/k|$) sees the Doppler-shifted IC wave
polarization as right- rather than left-circular, thus enabling gyro-resonance
\citep{tsurutani1997,terasawa2012}.
The resonance condition simplifies in the low-frequency limit, appropriate for
ICM plasmas with Alfv\'{e}n speed $\vA/c \ll 1$ and ion-electron mass ratio
$\mime \gg 1$:
\begin{equation} \label{eq:scaling}
    \frac{ p_{\prll} }{ m_\mt{e} c }
    \approx \frac{|\Omce|}{kc}
    \approx \frac{1}{k\rLi} \left(\frac{v_\mt{th,i}}{c}\right) \left(\frac{\mi}{\me}\right) .
\end{equation}
Here, $v_\mt{th,i} = \sqrt{3\kB T_\mt{i}/\mi}$ is ion thermal velocity.
The form of Eq.~\eqref{eq:scaling} anticipates that $k^{-1}$ is of order the
ion Larmor radius $\rLi$ for temperature-anisotropy-driven IC waves at marginal
stability \citep{davidson1975,yoon2010,sironi2015-i}.\footnote{
    For $k c/\ompi = \Delta/\sqrt{\Delta+1}$ at marginal stability
    \citep[Eq.~(6)]{davidson1975}, adopting
    $\Delta = S/{\beta_{\mt{i}\prll}}^{0.5}$ with order-unity constant $S$
    \citep{sironi2015-i} yields $k \rLi \approx S$ for $\Delta \ll 1$.
    Here $c/\ompi$ is ion skin depth and $\beta_{\mt{i}\prll}$ is
    $\vec{B}$-parallel ion beta.
}
For ICM temperatures $T_\mt{i} \approx T_\mt{e} \sim 1$--$10\unit{keV}$
\citep{chen2007}, IC waves with $k \rLi \sim 0.5$,
and $\mime=1836$ for a proton-electron plasma, we anticipate resonant momenta
\[
    p_{\prll} \sim 7\text{--}21 \me c ,
\]
within the expected range for fossil CRe in the ICM,
$p \sim 1\text{--}300 \me c$ \citep{pinzke2013}.
We thus expect that IC waves may efficiently scatter fossil CRe.

Gyroresonant IC wave scattering may energize CRe in at least two different
ways.
First, the non-zero phase velocity of IC waves will transfer energy from waves
to CRe via second-order Fermi acceleration \citep{fermi1949}, but this is slow
because the energy gain per cycle scales with the square of the scatterers'
velocity, $(\vA/c)^2 \ll 1$ for IC waves.
Second, pitch-angle scattering couples parallel and perpendicular momenta
$p_\prll$, $p_\perp$ and drives CRe towards isotropy.
Pitch-angle scattering, in isolation, conserves particle energy.
But, scattering during bulk deformation can heat particles via
\emph{magnetic pumping} if the scattering rate is comparable to the bulk
deformation rate \citep{berger1958,lichko2017}.

Magnetic pumping in a compressing plasma works as follows.
Because particle momenta $p_\perp$ and $p_\prll$ have different adiabatic
responses to compression, a scattering rate comparable to the bulk compression
rate can cause a net transfer of energy from $p_\perp$ to $p_\prll$ over one
compress-decompress cycle; this energy transfer may be linked to a phase
difference between pressure anisotropy and magnetic field compression
\citep{lichko2017}.
Magnetic pumping has been previously studied in the contexts of plasma
confinement, planetary magnetospheres, and the solar wind
\citep{alfven1950, schluter1957, berger1958, goertz1978,
borovsky1981, borovsky1986, borovsky2017, lichko2017, lichko2020, fowler2020}.

In high-$\bp$ plasmas with $\Delta > 0$, anisotropy-driven IC waves may not be
the dominant fluctuations.
Non-propagating structures created by the mirror instability are thought to
prevail over IC waves, based on theory \citep[e.g.,][]{shoji2009,isenberg2013}
and measurements in Earth's magnetosheath \citep{schwartz1996} and the solar
wind \citep{bale2009}.
Nevertheless:
IC waves may coexist with mirror structures;
IC waves appear in 3D hybrid simulations of turbulent high-$\bp$ plasma
\citep{markovskii2020,arzamasskiy2022};
there may be local regions of the ICM with reduced plasma $\bp$ or with reduced
electron/ion temperature ratio $\Te/\Ti$ \citep{fox1997} more conducive for IC
wave growth.
Mirror modes also have $k \sim {\rLi}^{-1}$, so they may non-resonantly scatter
fossil CRe and drive magnetic pumping as well.
The same will likely hold for firehose modes excited when $\Delta < 0$.

IC resonant scattering of relativistic MeV electrons also occurs in Earth's
radiation belts and can precipitate electrons into the upper atmosphere
\citep[e.g.,][]{thorne1971, meredith2003, zhang2016, adair2022}.
In particular, \citet{borovsky2017} studied the same mechanism as this
manuscript -- compression-driven IC waves energizing relativistic electrons via
magnetic pumping -- applied to Earth's outer radiation belt.

\section{Methods} \label{sec:methods}

We simulate continuously-compressed ICM plasma using the relativistic
particle-in-cell (PIC) code TRISTAN-MP \citep{buneman1993, spitkovsky2005}.
The PIC equations are solved in co-moving coordinates while subject to global
compression or expansion, as implemented by \citet{sironi2015-i}, similar to
hybrid expanding box simulations in the literature
\citep{liewer2001,hellinger2003-expand,hellinger2005,innocenti2019-code,bott2021}.
To do this, \citet{sironi2015-i} transform from the physical laboratory frame
$(t_\mt{lab},\vec{x}_\mt{lab})$ to a co-moving coordinate frame
$(t', \vec{x}')$ via a transformation law
$\vec{x}_\mt{lab} = \vec{L} \vec{x}'$, where:
\[
    \vec{L} 
    = \begin{pmatrix}
        a_x(t) & 0 & 0 \\
        0 & a_y(t) & 0 \\
        0 & 0 & a_z(t) \\
    \end{pmatrix} ,
\]
and the differential transformation law is:
\[
    \dtl\vec{x}_\mt{lab}
    = \vec{L} \dtl\vec{x}' + \dot{\vec{L}} \vec{x}' \dtl t' .
\]
The scale factors $a_x$, $a_y$, and $a_z$ are $>1$ for expansion and $<1$ for
contraction.
We report quantities (fields, particle positions, momenta, distribution
function moments) in physical CGS units in the plasma's local rest frame;
i.e., the unprimed coordinates $\dtl\vec{x}=\vec{L}\dtl\vec{x}'$ of
\citet{sironi2015-i}.

We use a 1D domain parallel to a background magnetic field $\vec{B}$, which
permits growth of parallel-propagating IC waves and precludes growth of the
mirror instability.
Our domain and magnetic field $\vec{B}$ are aligned along $y$;
all wavenumbers $k \equiv k_y$ in this manuscript.
We compress along both $x$ and $z$ axes by choosing scale factors:
\begin{equation} \label{eq:scale}
    a_x(t) = a_z(t) = \frac{1}{1 + q t}
\end{equation}
where $q > 0$ is a tunable constant controlling the compression rate.
We fix $a_y(t) = 1$.
The background field evolves consistent with flux freezing as
\[
    B_y = B_g(t) = B_0 (1+qt)^2
\]
where $B_0$ is the initial field strength.
The imposed $\vec{B}$-perpendicular compression conserves two particle
invariants, ${p_\perp}^2/B$ and $p_\prll$, if there is no wave-particle
interaction \citep[Appendix~A.2]{sironi2015-i}.

The ICM is modeled as a thermal ion-electron plasma with Maxwell-J\"{u}ttner
distributions of initial temperature $T_0$ and density $n_0$ for each species.
The fossil CRe are modeled as test particles, i.e. passive tracer particles,
which advance according to the electromagnetic fields on the grid but do not
contribute to the plasma dynamics---in the PIC algorithm, they have no weight
and so deposit no current.
The treatment of fossil CRe as passive tracers is motivated by their low
kinetic energy density, $\abt 10^4 \times$ smaller than the thermal ICM, in
cluster outskirts as simulated by \citet[Fig.~3]{pinzke2013}.
But, fossil CRe could become dynamically important in the recently-shocked ICM
responsible for radio relics; see, e.g., \citet[Fig.~11]{boess2022},
\citet{ha2022}.

Standard length- and time-scales are defined as follows for thermal plasma
species $s \in \{\mt{i}, \mt{e}\}$.
The signed, non-relativistic particle cyclotron frequency
$\Omega_\mt{s} = q_\mt{s} B/(m_\mt{s} c)$.
The plasma frequency
$\omega_\mt{ps} = \sqrt{4\pi n_\mt{s} e^2/m_\mt{s}}$.
The Larmor radius $\rho_\mt{s} = m_\mt{s} v_\mt{th,s} c/(e B)$, where
$v_\mt{th,s} = \sqrt{3\kB T_\mt{s}/m_\mt{s}}$ is a thermal velocity.
Subscript $0$ in $\Omega_\mt{s0}$, $\omega_\mt{ps0}$, $\rho_\mt{s0}$, and other
symbols hereafter means that the quantity is evaluated at $t=0$.
Subscripts $\perp$ and $\prll$ indicate vector projections with respect to the
background magnetic field direction $\hat{y}$.

Our results center on one ``fiducial'' simulation with
ion-to-electron mass ratio $\mime=8$,
initial plasma beta $\bpo = 16\pi n_0\kB T_0/{B_0}^2 = 20$,
initial Alfv\'{e}n speed $\vAo/c = B_0/\sqrt{4\pi(\mi+\me) n_0 c^2} = 0.067$
and compression timescale $q^{-1} = 800 {\Omcio}^{-1}$.
The choice of $\vAo/c$ is equivalent to choosing initial temperature
$\kB T_0/(\me c^2) = 0.2$ for fixed $\bpo$.
We use 16,384 particles per cell for the thermal plasma (i.e., 8,192 ions and
8,192 electrons per cell); Appendix~\ref{app:conv} shows convergence with
respect to the number of particles per cell.
The plasma skin depth $c/\sqrt{{\ompeo}^2+{\ompio}^2}$ is resolved with $5$
cells.
The domain size is $4608 \unit{\mt{cells}} = 307.2 c/\ompio = 79.3 \rLio$.
The Debye length $\lambda_\mt{De} = \sqrt{\kB T_0/(4\pi n_0 e^2)}$ is resolved
with 2.4 cells.
The numerical speed of light is $0.25$ grid cells per simulation timestep to
ensure that the Courant-Friedrichs-Lewy condition is satisfied for smaller
physical cell lengths at late simulation times
\citep[Appendix~A.1]{sironi2015-i}.
In each timestep, the electric current is smoothed with 32 passes of a
three-point binomial (``1-2-1'') filter, approximating a Gaussian filter with
standard deviation of $4$ cells \citep[Appendix C]{birdsall1991}.
Outputs are saved at $\abt 1 {\Omcio}^{-1}$ intervals.

We use two different initial test-particle CRe distributions $f(p) \mt{d}p$
depending on our analysis needs: $f(p)$ constant (flat) or $f(p) \propto
p^{-1}$ to uniformly sample $p$ or $\log p$ respectively.
Both distributions are isotropic.
The $f(p)$ constant case uses 2,880,000 CRe in $p=0$--$70\;\me c$, and the
$f(p)\propto p^{-1}$ case uses 14,400,000 CRe in $p=0.0014$ to $1400\;\me c$.
Neither case mimics nature, but the uniform $p$ and $\log p$ sampling means
that our results can be re-weighted to describe any initially isotropic CRe
distribution.
The test-particle distributions span the momentum range of CRe which should be
efficiently scattered by IC waves in our simulation:
$p_\prll \sim 4$--$25 \;\me c$ based on Eq.~\eqref{eq:scaling}.\footnote{
    Assuming $k\rLi \sim 0.5$ and $B_g(t)$ increasing $6\times$ from $t=0$
    to $1.5q^{-1}$.
}

Besides our fiducial simulation, we also run simulations with varying
$q$, $\mime$, $\vAo/c$, and $\bpo$;
detailed parameters are given in Appendix~\ref{app:param} and
Table~\ref{tab:param}.
The domain size is pinned to $\abt 80 \rLio$ for all such simulations.
The test-particle CRe spectrum is kept flat ($f(p)$ constant), but the upper
bound is re-scaled according to $(\mime)(v_\mt{th,i}/c)$ per
Eq.~\eqref{eq:scaling} to capture the momentum range of the expected IC
gyroresonance.
The simulations with varying $\bpo$ are not presented in the main text and
appear only in Appendix~\ref{app:offset}.
In cases with slow compression, e.g. $q^{-1}/\Omcio^{-1}=3200$ or
$\mi/\me=32$ in Table~\ref{tab:param}, we saw gyrophase-dependent numerical
errors in particle momenta when using single-precision (32 bit) floats in
the PIC algorithm.
We therefore use double-precision (64 bit) floats for all simulations in
the manuscript, except for convergence checks in Appendix~\ref{app:conv}.

Besides IC waves, whistlers (i.e., electron cyclotron waves) are excited by the
thermal electrons in our simulations.
To help separate the effects of whistler and IC waves upon fossil CRe energy
gain, we perform simulations in which one particle species, ions or electrons,
is compressed isotropically in order to suppress that species' cyclotron waves.
The species may still participate in plasma dynamics by generating currents.
To implement isotropic compression, we modify the co-moving momentum equation
(Boris particle push):
\begin{equation} \label{eq:boris}
    \frac{d\vec{p}}{dt}
    = - \dot{\vec{L}} \vec{L}^{-1} \vec{p}
    + q \left( \vec{E} + \frac{\vec{v}}{c} \times \vec{B} \right) .
\end{equation}
For the chosen species, we set the diagonal elements of
$\dot{\vec{L}} \vec{L}^{-1}$ in the Boris pusher to
$a_x=a_y=a_z = 1/(1+q_\mt{iso} t)$.
We choose $q_\mt{iso}=2q/3$ to match the initial energy input rate from
anisotropic compression; i.e., at $t=0$, the determinant
$\ell \equiv \det \vec{L} = 1/(1+q_\mt{iso}t)^3$ has first derivative equal to
that for the anisotropic $\ell = 1/(1+qt)^2$.
All other code in the PIC algorithm retains the anisotropic compression.
For electrons, isotropic forcing is only applied to regular particles (thermal
ICM) and not test particles (fossil CRe).

\section{Wave properties}

\subsection{Time evolution}

\begin{figure}
    \includegraphics[width=3.375in]{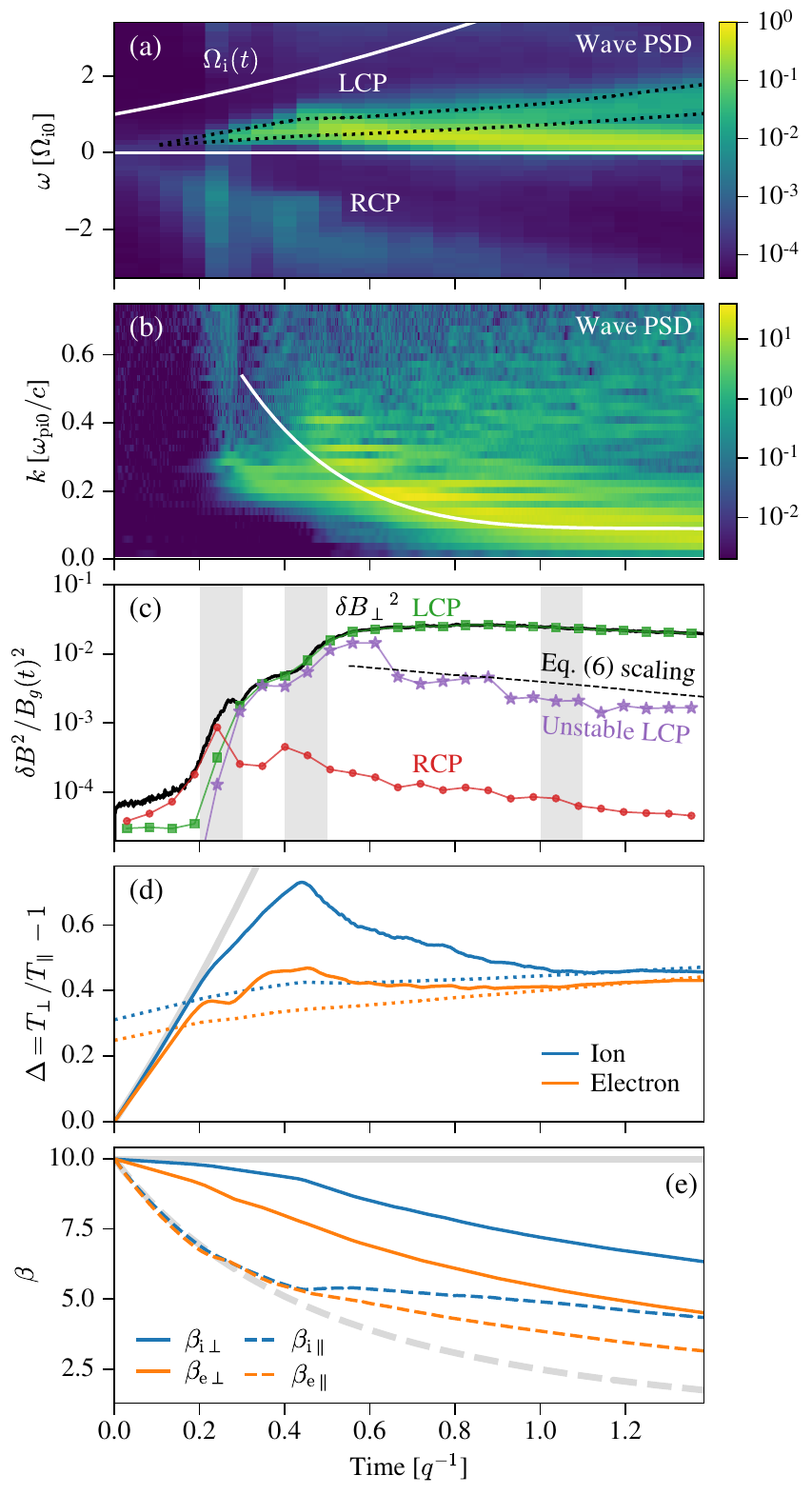}
    \caption{
        (a) Wave power spectral density (PSD) of $(B_z + i B_x)/B_g(t)$.
        Left- and right-circularly polarized (LCP, RCP) waves have
        $\omega > 0$, $< 0$ respectively.
        White line is ion cyclotron frequency $\Omci(t)$.
        Black dotted lines mark region of unstable IC waves with linear growth
        rate $\Gamma > 10^{-3} \Omcio$ from Eq.~\ref{eq:dispersion}.
        PSD is normalized so that $\omega$-axis average yields
        $\delta {B_\perp}^2/B_g(t)^2$.
        (b) Wave PSD with $k$ on the $y$-axis.
        Axis limits omit high-$k$ power to emphasize low-$k$ LCP waves.
        White curve is Eq.~\eqref{eq:kicdrift} for $t > 0.3 q^{-1}$.
        (c) Total magnetic fluctuation power $\delta {B_\perp}^2/B_g(t)^2$
        (solid black).
        Green squares, red circles respectively show LCP, RCP power from (a).
        Purple stars show power within unstable IC wave region from (a); i.e.,
        PSD between black dotted lines.
        Saturated wave scaling Eq.~\eqref{eq:driftkinetic2} (dashed black) is
        plotted with arbitrary normalization.
        Gray bands mark time intervals in
        Figs.~\ref{fig:fft}--\ref{fig:scattering}.
        (d) Ion (blue) and electron (orange) temperature anisotropy with
        best-fit scalings at marginal stability:
        $\Delta_\mt{i} = 1.00 {\beta_{\mt{i}\prll}}^{-0.5}$ (dotted blue),
        $\Delta_\mt{e} = 0.80 {\beta_{\mt{e}\prll}}^{-0.5}$ (dotted orange).
        (e) Ion and electron plasma beta perpendicular (solid) and parallel
        (dashed) to $\vec{B}$; colors as in (d).
        In (d-e), light gray curves are non-relativistic CGL predictions for
        adiabatic compression.
    }
    \label{fig:overview}
\end{figure}

The simulation evolves as follows.
The compression at first drives
$T_\perp \propto B_g(t) > T_\prll = \mt{constant}$ for all species while
conserving the adiabatic invariants of magnetized particles
\citep{northrop1963}, which can be recast in Chew-Goldberger-Low (CGL) fluid
theory as pressure or temperature invariants \citep{chew1956}.
Instability is triggered, and waves grow, between $t = 0.2 q^{-1}$ and
$0.5 q^{-1}$ (Fig.~\ref{fig:overview}(a-c)).
Right-circularly polarized (RCP) whistlers appear first and are the dominant
mode at $t=0.2 q^{-1}$, followed by left-circularly polarized (LCP) ion
cyclotron waves from $t=0.3$ to $0.5 q^{-1}$.
The wave polarizations are distinguished by Fourier transform of $B_z + i B_x$
in Fig.~\ref{fig:overview}(a), which separates LCP and RCP waves into
$\omega >0$ and $<0$ respectively, following \citet{ley2019}.
The wave fluctuation power $(\delta {B_\perp}/B_g)^2$ saturates at a
near-constant or slightly-decreasing level by $t\sim0.55q^{-1}$
(Fig.~\ref{fig:overview}(c));
while saturated, the IC wave power drifts towards lower $\omega$ and $k$
(Fig.~\ref{fig:overview}(a-b)).
We plot a manually-chosen approximation to the $k$-space drift,
\begin{equation} \label{eq:kicdrift}
    k_\mt{IC}(t) = \left[ 0.09 + 0.18 (1.5 - qt)^5 \right] \ompio/c ,
\end{equation}
in Fig.~\ref{fig:overview}(b), to be used later in this manuscript
(Sec.~\ref{sec:fpmodel}).

The saturated waves drive the ion and electron temperature anisotropy $\Delta$
away from CGL-invariant conservation and towards a marginally stable state at
late times $t \gtrsim 1 q^{-1}$ (Fig.~\ref{fig:overview}(d)).
At marginal stability, we expect $\Delta \propto \beta_{\mt{s}\prll}^{-0.5}$
for both ions \citep{gary1994-ic-sim,gary1994-ic-model,hellinger2006}
and electrons \citep{gary1996-whistler,gary2006-lec}, where
$\beta_{\mt{s}\prll} = 8\pi n_0\kB T_{\mt{s}\prll}(t)/{B_g(t)}^2$.
We fit the relation $\Delta \propto A_\mt{s} \beta_{\mt{s}\prll}^{-0.5}$
between $t = 1 q^{-1}$ and simulation's end to obtain
$A_\mt{i}=0.98 \pm 0.02$ and $A_\mt{e}=0.785 \pm 0.012$;
the best-fit relations are dotted lines in Fig.~\ref{fig:overview}(d).
The uncertainty on $A_\mt{i}$ and $A_\mt{e}$ is one standard deviation
estimated by assuming $\chi^2_\mt{reduced} = 1$, as no data uncertainty is used
in fitting.
We expect that the systematic uncertainty is larger.

When the IC waves saturate, we expect balance between compression increasing
$\Delta$ and wave pitch-angle scattering decreasing $\Delta$, as suggested by
the marginal-stability scaling in Fig.~\ref{fig:overview}(d).
This balance may be stated as:
\begin{equation} \label{eq:driftkinetic}
    \frac{\dtl \Delta}{\dtl t}
    = \frac{\dot{B}}{B} \left(\Delta + 1\right) - \nu \Delta (2\Delta + 3)
    \approx 0 ,
\end{equation}
which we obtain from moments of the Vlasov equation with a Lorentz-operator
scattering frequency $\nu$ constant with respect to momentum $p$ and
pitch-angle cosine $\mu \equiv p_\prll/p$ (Appendix~\ref{app:dk}), using a
drift-kinetic model as in \citet{zweibel2020,ley2022}
and following a similar argument as in \citet[Sec.~3.1.2]{kunz2020}.
If scattering scales like the quasi-linear approximation,
$\nu \propto \delta {B_\perp}^2$, then we expect
\begin{equation} \label{eq:driftkinetic2}
    \left(\frac{\delta B_\perp}{B_g(t)}\right)^2
    \propto \frac{\nu}{\Omci(t)}
    \approx \left(\frac{\dot{B}/B}{\Omci(t)}\right)
        \frac{\Delta_\mt{i} + 1}{\Delta_\mt{i}(2\Delta_\mt{i} +3)} \, .
\end{equation}
In taking $\Delta = \Delta_\mt{i}$, we assume that only ions source and control
the wave power $(\delta B_\perp/B)^2$ at late times.
In Fig.~\ref{fig:overview}(c), we show Eq.~\eqref{eq:driftkinetic2} computed
with arbitrary normalization and using
$\Delta_\mt{i} = T_{\mt{i}\perp}/T_{\mt{i}\prll}-1$ measured from the simulation.
Eq.~\eqref{eq:driftkinetic2} does not explain the total late-time wave power in
our simulation, but it better matches the power in currently-unstable IC waves
(Fig.~\ref{fig:overview}(a,c)).
We conjecture that waves in the unstable IC region may be most important for
regulating $\Delta$, in contrast to the stronger IC wave power at lower $k$.

The total plasma beta decreases to half its initial value by the simulation's
end, with ions hotter than electrons (Fig.~\ref{fig:overview}(e)).
At early times $t \lesssim 0.1 q^{-1}$, $\beta_{\mt{e}\perp}$ deviates from the
non-relativistic CGL prediction because electrons are almost relativistic with
$\kB T_\mathrm{0} = 0.2 m_\mt{e} c^2$.

\subsection{Wave identification} \label{sec:wave-id}

\begin{figure}
    \plotone{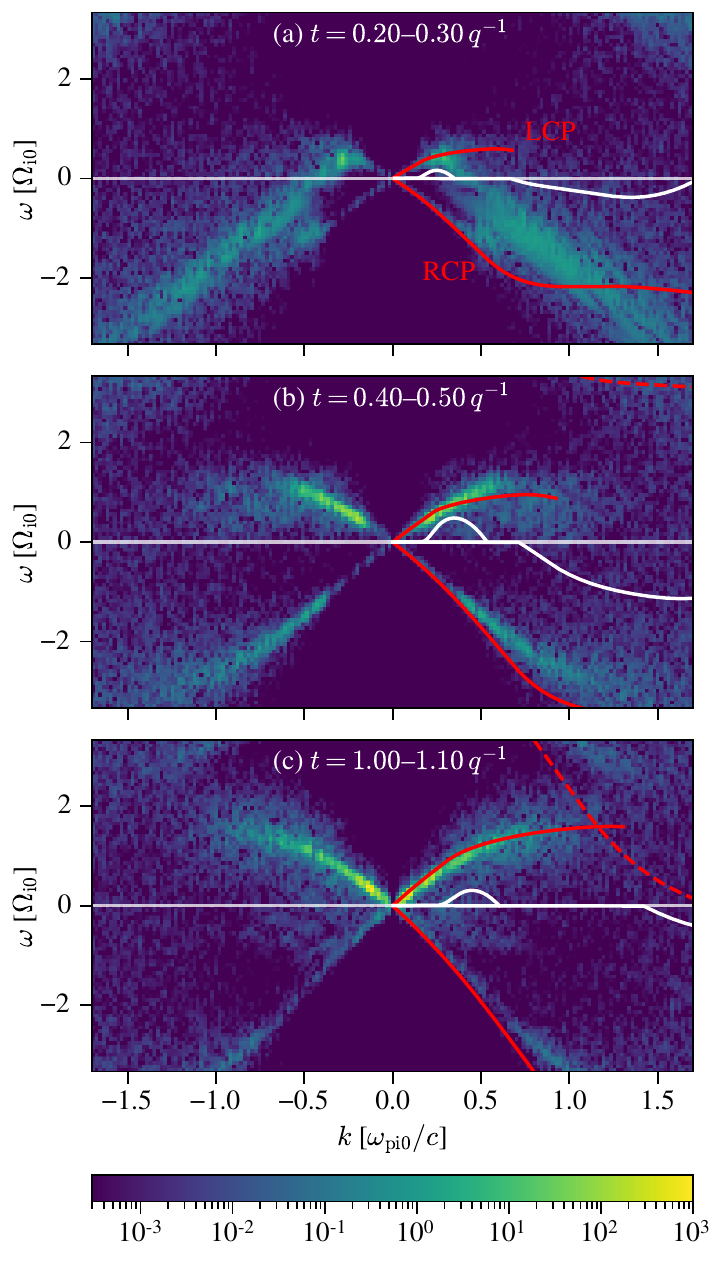}
    \caption{
        Power spectrum of $B_z + i B_x$ for three different time intervals.
        Positive/negative $\omega$ are LCP/RCP respectively.
        Real and imaginary parts of the bi-Maxwellian dispersion relation,
        $\omega(k)$ and $\Gamma(k)$ computed from Eq.~\eqref{eq:dispersion},
        are red and white curves respectively.
        Only $\Gamma > 0$, indicating instability, is shown.
        The IC and whistler growth rates are plotted $5\times$ and $2\times$
        larger than their true values for visibility.
        Dashed red curve is high-frequency whistler mode, aliased due to finite
        time sampling.
        The dispersion curves truncate when damping becomes strong,
        $\Gamma(k) < -|\omega(k)|$.
    }
    \label{fig:fft}
\end{figure}

Let us now more closely study wave properties and evolution.
To predict wave $\omega$, $k$, and damping/growth as a function of time, we
solve the non-relativistic dispersion relation for $\vec{B}$-parallel
electromagnetic waves in a bi-Maxwellian ion-electron plasma:
\begin{align}
    \label{eq:dispersion}
    &D^\pm = 1 - \frac{k^2 c^2}{\omega^2} \\
    &+ \sum_\mt{s} \left(\frac{\omega_\mt{ps}}{\omega}\right)^2
    \left[
        \zeta_0 Z\left(\zeta_{\pm1}\right)
        - \frac{1}{2} \left( \frac{T_\mt{s\perp}}{T_\mt{s\prll}} - 1 \right)
        Z'\left(\zeta_{\pm1}\right)
    \right]
    \nonumber
\end{align}
as stated in \citet{davidson1975} and \citet[Sec.~11-2]{stix1992}, keeping only
the $n=0,\pm1$ resonant terms.
The subscript $\mt{s}=\mt{i},\mt{e}$ indexes component species,
$Z(\zeta) = 2 i \exp(-\zeta^2) \int_{-\infty}^{i\zeta} \exp(-t^2) \dtl t$
is the plasma dispersion function \citep{fried1961},
$\zeta_{n} = (\omega - n\Omega_\mt{s})/k_\prll w_{\mt{s}\prll}$,
and $w_{\mt{s}\prll} = \sqrt{2 \kB T_{\mt{s}\prll}/m_s}$.
We approximate $T_{\mt{s}\prll}$ and $T_{\mt{s}\perp}$ using the second moments
of both ion and electron distributions in our simulations.
In Eq.~\ref{eq:dispersion}, $\omega = \omega_\mt{R} + i \Gamma$ is complex, but
in all other text and figures, $\omega$ refers only to the real angular
frequency $\omega_\mt{R}$ unless otherwise noted.
The imaginary part $\Gamma > 0$ for instability and $< 0$ for damping.
We use Eq.~\ref{eq:dispersion} to show the unstable $\omega$ range for LCP
waves over time in Fig.~\ref{fig:overview}(b), and to show the expected
$\omega$ and $k$ for both LCP and RCP waves in the $\omega$-$k$ power spectra
of Fig.~\ref{fig:fft}.

We note several features of interest in the $B_z + i B_x$ spectrogram
(Fig.~\ref{fig:overview}(b)).
LCP and RCP modes both appear at $t \sim 0.2 q^{-1}$.
The LCP mode is more monochromatic and has lower $\omega$, while the RCP mode
has broader bandwidth and higher $\omega$.
The LCP modes persist from $t > 0.2 q^{-1}$ through the rest of the simulation.
The RCP modes appear in two transient bursts, at $t=0.2$ and $0.4 \;q^{-1}$,
and the second RCP burst coincides with a growth of LCP power and near-peak ion
anisotropy $\Delta_\mt{i}$.
Some RCP power aliases from $\omega < 0$ into $\omega > 0$ at the top of
Fig.~\ref{fig:overview}(b) and in each panel of Fig.~\ref{fig:fft}.

The LCP power splits into high- and low-frequency bands at
$t \approx 0.8$--$1.0 q^{-1}$ (Fig.~\ref{fig:overview}(a)); each band continues
to respectively rise and fall in frequency over time.
The high-frequency LCP power lies within the expected $\omega$ range of IC wave
instability as predicted by Eq.~\ref{eq:dispersion}.
The low-frequency LCP power resides in a frequency/wavenumber range that is not
expected to spontaneously grow IC waves.
We remain agnostic about why the low-frequency LCP power evolves towards low
$k$, but we note that \citet[Fig.~9]{ley2019} saw a similar drift of IC wave
power to low $k$ in a shearing-box PIC simulation.
In Appendix~\ref{app:halt-compress}, we show that wave power drifts to low
frequencies even if compression halts at $t=0.5q^{-1}$, so the
low-frequency power drift is not caused by external compression or by a
numerical artifact of the comoving PIC domain.

We verify that LCP and RCP modes are IC waves and whistlers respectively by
inspecting $\omega$--$k$ power spectra in three time intervals
(Fig.~\ref{fig:fft}).
The LCP wave power agrees well with the predicted $(\omega,k)$ from
Eq.~\eqref{eq:dispersion} in all time snapshots of Fig.~\ref{fig:fft},
and the previously-noted high-frequency band in Fig.~\ref{fig:overview}(b)
agrees well with the prediction for IC wave instability.
The RCP wave power agrees with the bi-Maxwellian whistler dispersion in some
respects.
The phase speed $\omega/k$ agrees with Eq.~\eqref{eq:dispersion} at later times
(Fig.~\ref{fig:fft}(b-c)).
In simulations with higher $\mime$ (Appendix~\ref{app:offset}), the RCP phase
speed $\omega/k$ increases with respect to the LCP phase speed and continues to
agree with Eq.~\eqref{eq:dispersion}.
But, the RCP wave power disagrees with the bi-Maxwellian dispersion curve in
some respects.
At early times $t=0.2$--$0.3 q^{-1}$, the RCP mode is offset towards higher
$k$ than expected for the whistler mode; it does not appear to lie on a curve
passing through $(\omega,k)=(0,0)$.
At later times, the RCP power shows better agreement with the whistler mode:
the $k$ offset disappears and RCP power connects continuously to
$(\omega,k)=(0,0)$ (Fig.~\ref{fig:fft}(b-c)).
The later-time RCP power also has $\omega$ somewhat lower than predicted by
Eq.~\eqref{eq:dispersion} for $k = 0.5$--$1.0\ompio/c$
(Fig.~\ref{fig:fft}(b-c)).
Some more observations on the RCP mode are in Appendix~\ref{app:offset}.
All considered, despite the imperfect agreement with Eq.~\eqref{eq:dispersion},
we attribute RCP waves to thermal electron anisotropy and call them whistlers
hereafter.

Eq.~\eqref{eq:dispersion} is approximate, as particles are not exactly
bi-Maxwellian.
Wave scattering alters distributions to quench instability, and the
resulting anisotropic distributions can be stable to ion cyclotron waves
\citep{isenberg2013}.
Appendix~\ref{app:halt-compress} checks the frequency of waves driven
unstable by the actual particle distribution, and we find that the
resulting waves do lie in a high-frequency LCP power band as predicted by
Eq.~\eqref{eq:dispersion}, validating our use of the bi-Maxwellian
approximation in this context.

Eq.~\eqref{eq:dispersion} also does not account for the background plasma
density and magnetic field varying during instability growth; the plasma
properties are assumed to vary on a much longer timescale than is relevant
to the linear dispersion calculation.
The maximum IC growth rate predicted by Eq.~\eqref{eq:dispersion} is
$\abt 0.1 {\Omcio}^{-1}$ at $t\approx0.5q^{-1}$ (Fig.~\ref{fig:fft}(b)),
which is $80\times$ faster than the compression rate $q$.
The growth rate may be smaller in practice due to particles quenching their
own instability; nevertheless, we expect that waves should grow on a short
timescale that's well separated from the compression time.

\section{Wave scattering} \label{sec:scatt}

\begin{figure*}
    \plotone{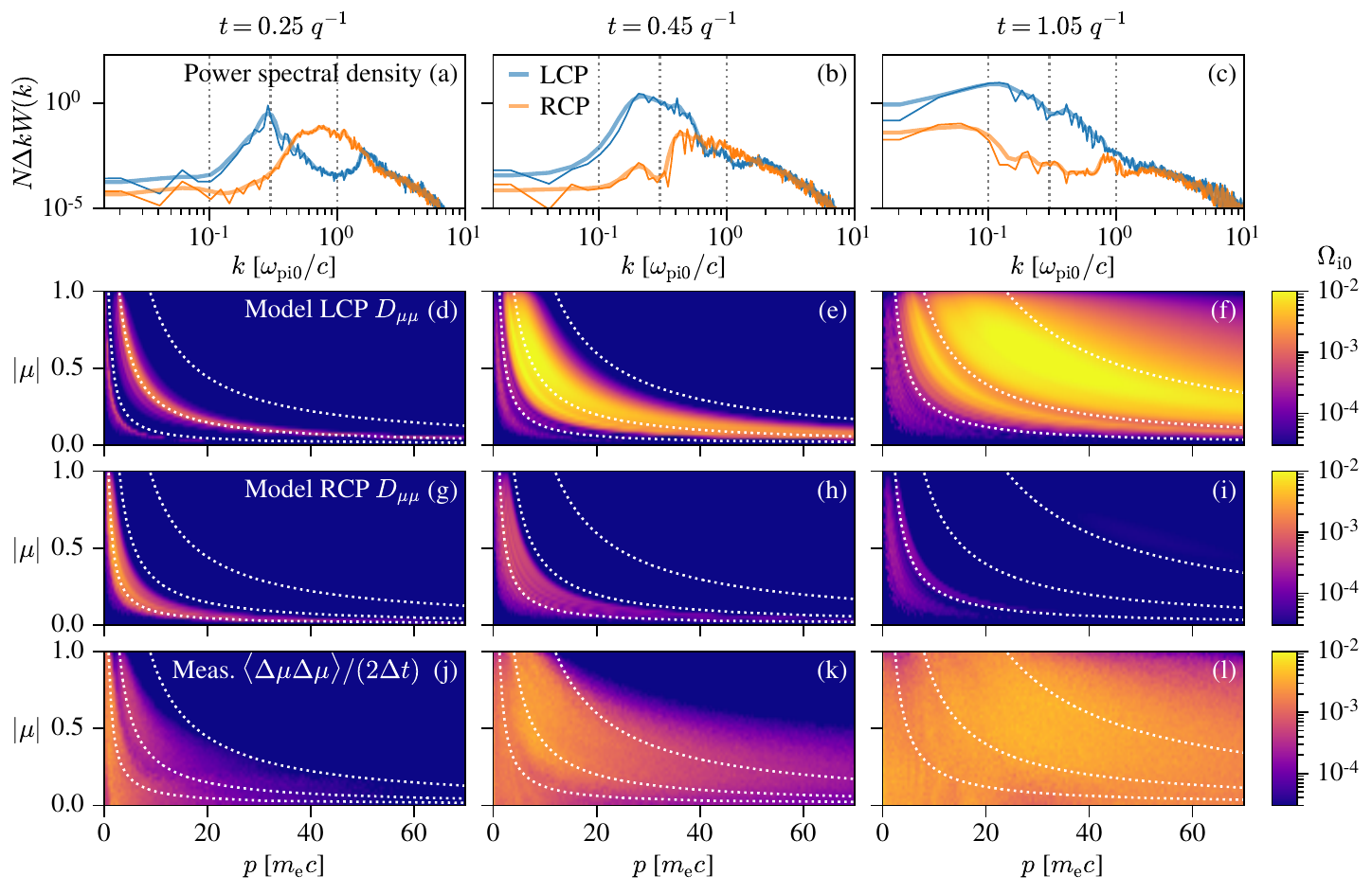}
    \caption{
        Measured scattering compared to quasi-linear model at three different
        times advancing from left to right.
        (a-c): LCP (blue line) and RCP (orange line) magnetic power spectra
        $W_\mt{L}$ and $W_\mt{R}$.
        Spectra are normalized using the number of grid points $N$ and the
        Fourier spacing $\Delta k$ so that
        $\int (W_\mt{L} + W_\mt{R}) \dtl k = \left(\delta B_\perp/B_g\right)^2$.
        Thick line is spectrum smoothed with Hanning window of length
        $0.14 \ompio/c$ (7 points), as used to compute $D_{\mu\mu}$ in panels
        (d-i) below.
        (d-f): Quasi-linear model for LCP (IC) wave diffusion computed using
        $W_\mt{L}$.
        (g-i): Quasi-linear model for RCP (whistler) wave diffusion computed
        using $W_\mt{R}$.
        (j-l): Measured pitch-angle scattering rate
        $\langle\Delta\mu\Delta\mu\rangle/(2\Delta t)$ for test-particle CRe.
        White dotted lines mark particles resonant with wavenumbers, from left
        to right: $k_\mt{res} c/\ompio = 1$, $0.3$, and $0.1$, according to
        Eq.~\eqref{eq:qlt-kres}; same wavenumbers are marked as vertical dotted
        black lines in panels (a-c).
        All scattering rates in panels (d-i) are scaled by $\Omcio$.
    }
    \label{fig:scattering}
\end{figure*}

Let us compare the CRe scattering directly measured in our simulations against
the quasi-linear theory (QLT) description of resonant scattering as a diffusive
process, in the limit of weak, uncorrelated, and broad-band waves
\citep{kennel1966-petschek,kennel1966-engelmann,jokipii1966,kulsrud1969}.
In particular, we wish to check the following.
(1) Do particles with $90^\circ$ pitch angle (i.e., $p_\perp \gg p_\prll$)
scatter efficiently in our simulations?  As $p_\prll \to 0$, the resonant
wavenumber $k_\mt{res} \to\infty$, and particles cannot scatter at
exactly $90^\circ$ pitch angle in QLT.
(2) Does the resonant QLT description hold for our simulations?
The saturated wave power $\delta B_\perp/B \sim 0.1$
(Fig.~\ref{fig:overview}(a)) may be too strong to satisfy QLT \citep{liu2010}.
Strong waves may lead to, for example, momentum-space advection instead of
diffusion \citep{albert2009}.

We compute the QLT diffusion coefficient $D_{\mu\mu}$ for pitch-angle cosine
$\mu = \cos\alpha = p_\prll/p$, assuming low-frequency ($\omega \approx 0$)
waves, following \citet{summers2005}:
\begin{equation} \label{eq:qlt}
    D_{\mu\mu}(p,\mu)
        \approx (1-\mu^2) \frac{\pi}{2} \frac{|\Omce(t)|}{\gamma}
            \frac{k_\mt{res} W(k_\mt{res})}{{B_g(t)}^2/(8\pi)} \, .
\end{equation}
Because momentum scattering is subdominant in our simulations
(Sec.~\ref{sec:vAc-mime}), and is expected to be even more subdominant for
lower $\vA/c$ in the real ICM, we neglect the QLT diffusion coefficients
$D_{pp}$ and $D_{p\mu}$ for now.
The resonant signed wavenumber
\begin{equation} \label{eq:qlt-kres}
    k_\mt{res} = \pm \frac{e B_g(t)}{\mu p c} \, ,
\end{equation}
with $+$ and $-$ signs for CRe resonance with IC and whistler waves
respectively.
We take $W(k)$ to be the \emph{two-sided} wave power spectrum of
$\delta B_\perp^2$ measured directly from our simulation, with the sign of $k$
specifying propagation direction.
We decompose $W(k) = W_\mt{L}(k) + W_\mt{R}(k)$ into LCP and RCP pieces by
Fourier transforming $B_z + i B_x$ over a time window of length
$18.9{\Omcio}^{-1}$, which is $4\times$ larger than the timestep used to
measure particle scattering.
Power at $\omega \ge 0$ is assigned to $W_\mt{L}$ and the remainder to
$W_\mt{R}$.
We smooth $W_\mt{R}(k)$ and $W_\mt{L}(k)$ with a Hanning window of length
$0.14 \ompio/c$ (7 points) and then linearly interpolate to compute
$D_{\mu\mu}$ for arbitrary $(p,\mu)$.
Because our simulation has balanced forward- and backward-propagating waves, we
average $D_{\mu\mu}$ over $\mu < 0$ and $\mu > 0$ in Fig.~\ref{fig:scattering}.

We directly measure $\langle \Delta\mu \Delta\mu \rangle / (2 \Delta t)$ by
computing $\Delta\mu = \mu(t+\Delta t) - \mu(t)$ over an output timestep
$\Delta t = 4.7 {\Omcio}^{-1}$ for each test-particle CRe.
The pitch angle $\alpha$ is defined with respect to the background field
$B_g(t) \hat{y}$.
Then, we compute particle-averaged $\langle \Delta \mu \Delta \mu \rangle$ as a
function of phase-space coordinates $(\mu,p)$ using 50 bins over
$|\mu| \in [0,1]$ and 140 bins over $p \in [0,70] \me c$.
The choice of $\Delta t$ affects the shape and strength of scattering regions
in Fig.~\ref{fig:scattering}(j-l).
We find that timesteps $\Delta t = 4.7$--$18.8\; {\Omcio}^{-1}$ give somewhat
consistent scattering region shapes, but shorter timesteps
$\Delta t = 0.9$--$1.9\; {\Omcio}^{-1}$ do not resolve the scattering
interaction, especially for the highest $p$ CRe.
Appendix~\ref{app:idt} further shows and discusses the effect of varying
$\Delta t$ in our scattering measurement.

Fig.~\ref{fig:scattering} compares the measured pitch-angle scattering rates
$\langle \Delta \mu \Delta \mu \rangle / (2\Delta t)$
(Fig.~\ref{fig:scattering}(j-l)) to the predicted rates $D_{\mu\mu}$ from LCP
(Fig.~\ref{fig:scattering}(d-f)) and RCP (Fig.~\ref{fig:scattering}(g-i)) waves
at $t = 0.25$, $0.45$, and $1.05 q^{-1}$.
The smoothed $W_\mt{L}$ and $W_\mt{R}$ used to compute $D_{\mu\mu}$ are shown
in Fig.~\ref{fig:scattering}(a-c); the one-sided spectra, as normalized, are
averages of two-sided spectra over $k >0$ and $k < 0$.
The full QLT prediction for $D_{\mu\mu}$ is the sum of the middle two rows
(d-i), which separate the ion cyclotron and whister contributions to show their
relative importance.
White dotted lines mark all particles resonant with a wave of given
$k_\mt{res}$ according to Eq.~\eqref{eq:qlt-kres}.
At $t=0.25 q^{-1}$ (left column), whistler power is strong and the particles
most efficiently scattered have small momenta $p \sim 1$--$5 \;\me c$.
At $t=0.45 q^{-1}$ (middle column), ion cyclotron power has overtaken whistlers
in strength, with most resonant scattering predicted at the $k=0.3 \ompio/c$
contour, though the measured scattering $\langle\Delta\mu\Delta\mu\rangle$ has
broader bandwidth in $(\mu,p)$ space and does not exactly follow the resonant
contour shape of Eq.~\eqref{eq:qlt-kres}.
At $t=1.05 q^{-1}$ (right column), the wave power is saturated
(Fig.~\ref{fig:overview}(a)) and the IC spectrum has broadened to
$k=0.1 \;\ompio/c$, seen in both the 1D wave spectrum (top row) and the QLT
prediction (second row).

As time progresses, both the measured and modeled scattering extend towards
larger $p$ due to two effects.
First, the increase in $B_g(t)$ leads to rightward drift of the resonant
contours $p \propto B_g(t)/\mu$ (Eq.~\eqref{eq:qlt-kres}) for fixed
$k_\mt{res}$.
Second, the saturated wave power drifts towards smaller $k$ over time
(Fig.~\ref{fig:overview}(c), Fig.~\ref{fig:scattering}(b-c)).
Comparing Fig.~\ref{fig:scattering}(e) and (f), the QLT-predicted scattering
expands from the $k=0.3\;\ompio/c$ contour to $k=0.1\;\ompio/c$ as time
progresses.
Likewise, comparing Fig.~\ref{fig:scattering}(k) and (l), the measured
scattering expands beyond the $k=0.1\;\ompio/c$.
The drift of $k$-resonant surfaces through momentum space due to both effects
allows the cyclotron modes to interact with and scatter a larger volume of CRe
than would otherwise be possible.

The measured scattering differs from QLT in some respects.
The scattering region in $(p,\mu)$ is continuous through the $\mu=0$
($\alpha=90^\circ$) barrier, and the region is more extended in $(p,\mu)$ space
than the QLT prediction.
Scattering through $\mu=0$ may be explained by mirroring of particles with
$\mu^2 < (\delta B_\perp/B_g)_\mt{m}^2 / 2$ \citep[Eq.~(22)]{felice2001}, where
$(\delta B_\perp/B_g)_\mt{m}^2$ is power at the specific wavenumber(s)
responsible for non-resonant mirroring.
The total wave power (Fig.~\ref{fig:overview}(a)) sets an upper bound
$(\delta B_\perp/B_g)_\mt{m}^2 \lesssim 0.03$, and so we expect mirroring to be
important at $|\mu| \lesssim 0.12$.
We speculate that resonance broadening \citep[e.g.,][]{tonoian2022} or a
non-magnetostatic calculation with $\omega/k \ne 0$ may also expand the
scattering extent in $(p,|\mu|)$.
In particular, the magnetostatic assumption is less valid for the higher
$\vA/c$ in our simulations as compared to real ICM.
See also \citet{holcomb2019} for further recent discussion.

\section{Particle spectrum from magnetic pumping} \label{sec:spectrum}

\begin{figure*}
    \plotone{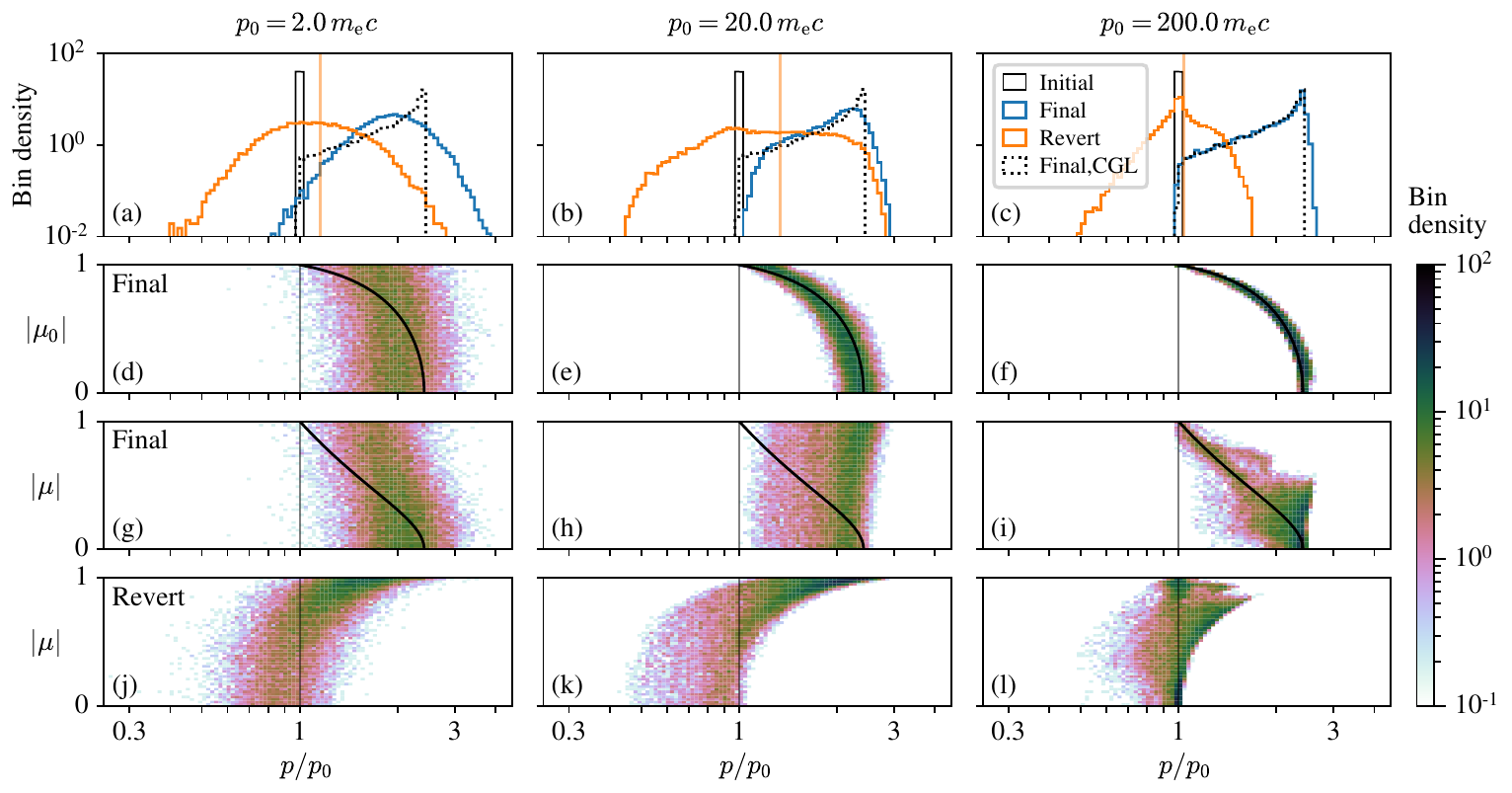}
    \caption{
        CRe response to compression and adiabatic decompression (``revert'')
        for three narrow initial distributions of mean momentum
        $p_0 = 2, 20, 200 \me c$ (left to right), and width $\pm 0.01 p_0$.
        (a-c): 1D CRe spectrum in $p$ at $t=0$ (``Initial'', black), end of
        simulation (``Final'', blue), and after adiabatic decompression
        (``Revert'', orange).
        For comparison, the CRe spectrum that would result from pure adiabatic
        compression is ``Final,CGL'' (dotted black).
        Mean momentum after compress-revert cycle is vertical orange line.
        (d-f): CRe $(p,|\mu_0|)$ distribution shows that the energy gain becomes
        closer to adiabatic, and correlates more with $|\mu_0|$, as $p_0$
        increases.
        (g-i): Final, compressed CRe $(p,|\mu|)$ distribution at end of
        simulation.
        (j-l): Revert CRe $(p,|\mu|)$ distribution, after a compress-revert
        cycle.
        In panels (d-l), vertical black line indicates starting momentum, and
        thick black curve marks shape of expected particle distribution for
        purely adiabatic compression.
        Histogram bin densities are counts divided by both 1D/2D bin size and
        total count of particle sample.
    }
    \label{fig:spectrum}
\end{figure*}

We now seek a time-integrated view of energy gain due to magnetic pumping from
IC wave scattering during compression.
Some particles scatter more efficiently and at different times than others, and
it follows that some fossil CRe may gain more energy from magnetic pumping than
others.

To frame the problem, we ask: given CRe of initial momentum $p_0$ at $t=0$,
what is their energy gain due to magnetic pumping during compression?
We consider the following hypothetical scenario.
After a compression to time $t$ in our simulation, let the test-particle CRe
decompress back to their initial volume, \emph{with no further wave
scattering during decompression};
i.e., map $p_\perp \to p_\perp \left[B_0/B(t)\right]^{1/2}$ and hold $p_\prll$
constant for all particles.
We call this adiabatic decompression a ``reversion'' of the particle
distribution, and we say that the particles have undergone a
``compress-revert'' cycle.
The decompressed particle energy is defined as
\[
    \gamma_\mt{revert}(t)
    = \sqrt{1 + {p_\prll(t)}^2 + {p_\perp(t)}^2 \left[B_0/B(t)\right] } .
\]
One cycle of compression to arbitrary time $t$, followed by a revert, yields an
energy gain:
\[
    \Delta U_\mt{revert}
    = U_\mt{revert} - U_0
    = \langle \gamma_\mt{revert}(t) \rangle - \langle \gamma(t=0) \rangle
\]
where $U(t) = \langle \gamma(t) \rangle$, $U_0 = U(t=0)$, and angle brackets
$\langle\cdots\rangle$ are ensemble averages over particles in an initial
momentum bin $p_0$.
Recall that our initial test-particle CRe distribution is isotropic; i.e.,
uniform on $\mu \in [-1,+1]$.
We use $\Delta U_\mt{revert}(t)$ as a proxy for magnetic pumping efficiency.

The ``revert'' is artificial; particles may scatter during decompression.
But, the compress-revert cycle permits us to focus solely on magnetic pumping
due to compression-driven waves, without needing to also study and separate the
effect of decompression-driven waves (e.g., firehose).

\begin{figure}
    \includegraphics[width=3.375in]{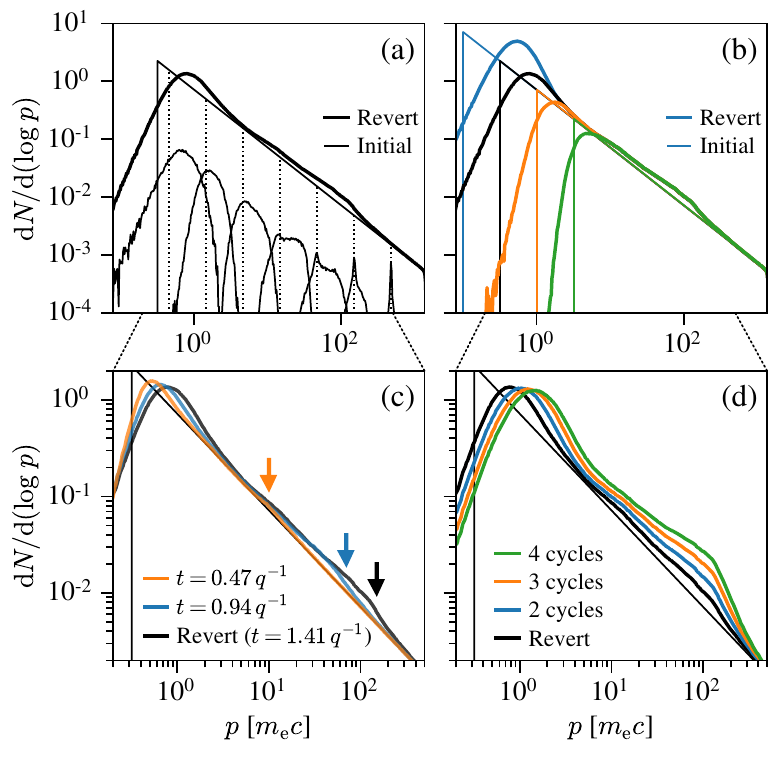}
    \caption{
        Convolution of an initial CRe spectrum $f(p) \dtl p = p^{-2} \dtl p$
        with the Dirac-delta spectrum response $G(p',p)$ to one and multiple
        compress-revert cycles.
        (a) Initial spectrum $f(p)$ (thin black) and revert spectrum after an
        assumed adiabatic decompression (thick black).
        Seven kernels $G(p',p)$ that contribute to the revert spectrum are
        shown, with their initial $p_0$ shown by dotted vertical black lines.
        (b) The low-$p$ bound on $f(p)$ sets the amplitude and position of a
        thermal bump at $p \sim 1\;\me c$, but does not strongly alter the
        ion cyclotron (IC) resonance bump at $p \sim 10$--$100 \;\me c$.
        We show four different bounds: $p = 10^{-1}$ (blue), $10^{-0.5}$
        (black), $10^0$ (orange), and $10^{0.5} \;\me c$ (green).
        (c) Compression duration determines the maximum $p$ attained by the IC
        bump.
        Here we show $f_\mt{revert}$ from our fiducial spectrum computed at
        three different times: $t=0.47 q^{-1}$ (orange), $t=0.94 q^{-1}$
        (blue), and $t=1.41 q^{-1}$ (black).
        The black curve is computed at simulation's end.
        Corresponding colored arrows indicate the right-ward extent of the IC
        bump in $p$.
        (d) Effect of multiple revert cycles, assuming that the spectrum is
        instantly re-isotropized after each revert cycle.
        In all of panels (a-d), the black curve $f_\mt{revert}$ is the same.
    }
    \label{fig:spectrum-convolve}
\end{figure}

We shall now seek to understand how particles respond to a compress-revert
cycle, before proceeding to use $\Delta U_\mt{revert}$ as a proxy for magnetic
pumping efficiency.
In Figs.~\ref{fig:spectrum}--\ref{fig:spectrum-convolve}, we use a
test-particle CRe spectrum $dN/dp = f(p) \propto p^{-1}$ that uniformly samples
$\log p$ with $p \in[0.0014, 1400] \me c$ using 14,400,000 particles.
But, we re-iterate that our results can be re-weighted to apply to any
initial $f(p)$, and Fig.~\ref{fig:spectrum-convolve} shows one such
re-weighting to $f(p) \propto p^{-2}$.

Fig.~\ref{fig:spectrum} shows one compress-revert cycle acting upon the
simulated CRe, where the ``Final'' particle distribution is from the
simulation's end, and the ``Revert'' particle distribution is taken after one
compress-revert cycle.
The ``Final,CGL'' distribution shows the same compression as for ``Final'',
but without scattering.
We call attention to four points.
First, the ``Revert'' particle spectrum is skewed; although the mean ``revert''
particle momentum is $\abt 1.1$--$1.3\times p_0$, individual particles may
be energized up to $\abt 2.4\times p_0$ (Fig.~\ref{fig:spectrum}(a-c)).
Second, scattering is strongest for low starting $p_0$ and weakens towards
higher $p_0$, as judged by the particles' deviation from the predictions for
adiabatic compression and adiabatic decompression
(Fig.~\ref{fig:spectrum}(d-l), black curves).
Third, the final particle momentum correlates with the cosine of the particle's
initial pitch angle $\mu_0$, and that correlation strengthens for larger
$p_0$ (Fig.~\ref{fig:spectrum}(d-f)).
The energy gain for particles with large $p_0$ is nearly consistent with
adiabatic compression, shown by comparing the ``Final'' particle distributions
to the ``Final,CGL'' curve in Fig.~\ref{fig:spectrum}(a-c) and thick black
curves in Fig.~\ref{fig:spectrum}(d-l).
Fourth, the``Final'' particle distribution extends rightwards of the expected
maximum momentum from adiabatic compression alone, $p_0 \sqrt{B(t)/B_0}$,
from comparing ``Final'' and ``Final,CGL'' distributions in
Fig.~\ref{fig:spectrum}(a-c).
We attribute the particles with $p > p_0 \sqrt{B(t)/B_0}$ to momentum diffusion
$D_{pp}$; the number of such particles decreases as we lower $\vAo/c$ towards
realistic values for the ICM and hence decrease $D_{pp}$.

We can model the magnetic pumping upon any isotropic CRe spectrum $f(p)$ by
computing the response of a Dirac delta distribution $\delta(p-p_0)$ to one
compress-revert cycle, for multiple choices of constant $p_0$, in the spirit of
a Green's function.
Let $p$ and $p'$ be momentum coordinates before and after a compress-revert
cycle respectively.
Define $G(p',p_0) \dtl p'$ to be the distribution obtained by applying one
compress-revert cycle to an initial distribution
$f(p) \dtl p = \delta(p-p_0) \dtl p$ with $p_0$ an arbitrary constant, similar
to Fig.~\ref{fig:spectrum}(a-c).
To construct $G$, we average over $\mu$, even though the particle spectrum
after a compress-revert cycle is not isotropic (Fig.~\ref{fig:spectrum}(j-l)).
Then, the action of one revert cycle upon $f(p)$ is:
\begin{equation} \label{eq:spectrum-conv}
    f_\mt{revert}(p') = \int f(p) G(p',p) \dtl p
\end{equation}
for any $f(p)$.
To implement Eq.~\eqref{eq:spectrum-conv} numerically, we compute $G(p',p)$
for each of 300 logarithmically-spaced bins over $p \in [0.0014, 1400] \me c$
with 96,000 test-particle CRe per bin.

Fig.~\ref{fig:spectrum-convolve} demonstrates the effect of magnetic pumping
for an ``Initial'' spectrum $f(p) \dtl p \propto p^{-2} \dtl p$ with lower
bound $p = 10^{-0.5} \me c$.
The ``Revert'' spectrum $f_\mt{revert}$ has two distinct bumps compared to the
Initial spectrum (Fig.~\ref{fig:spectrum-convolve}(a)).
We attribute the higher-$p$ bump at $p \sim 10$--$100\;\me c$ to the IC wave
resonance; hereafter, we call this the ``IC bump``.
The lower-$p$ bump with maximum at $p \sim 1 \;\me c$ has shape similar to a
thermal Maxwell-J\"{u}ttner distribution.
At high energies $p \gtrsim 300 \;\me c$, particle momenta remain nearly
adiabatic through a compress-revert cycle, as previously seen in
Fig.~\ref{fig:spectrum}(c,f,i,l).
We visualize the convolution of $f(p)$ by plotting the kernels $G(p',p)$ for
various $p$ (Fig.~\ref{fig:spectrum-convolve}(a)); these kernels are
constructed using the same procedure as
the 1D ``Revert'' spectra in Fig.~\ref{fig:spectrum}(a-c), up to
details of numerical binning and normalization.

The IC bump in $f_\mt{revert}(p)$ has an upper bound at $p \sim 100 \;\me c$
that is not exceeded by multiple pump cycles.
What sets this $p$ bound?
We attribute this bound to the rightward skew of the convolution kernel
$G(p',p)$, most visible for the kernels with $p$ between $10^1$ and
$10^2 \;\me c$ in Fig.~\ref{fig:spectrum-convolve}(a).
In contrast, the mean ($\mu$-averaged) energy gain after one compress-revert
cycle has a maximum of $\abt 30\%$ for CRe with initial momenta
$p_0 \sim 20$--$30 \;\me c$, which we will shortly see in
Fig.~\ref{fig:energygain}; see also the mean energy gain (vertical orange
lines) in Fig.~\ref{fig:spectrum}(a-c).
A mean energy gain of $1.3\times 30\;\me c$ does not easily explain the
increase in $f_\mt{revert}(p)$ at $p \sim 100 \;\me c$.

Is the IC bump in $f_\mt{revert}(p)$ sensitive to our choice of the
low-$p$ boundary for $f(p)$?
Fig.~\ref{fig:spectrum-convolve}(b) shows that altering the low-$p$ cut-off on
$f(p)$ also alters the amplitude and peak momentum of the thermal bump;
i.e., all electrons below $p \sim 1\;\me c$ are re-organized into a thermal
distribution.
Lowering the $p$ boundary of our input spectrum places more electrons into this
thermal bump.
The IC bump is not affected by the low-$p$ boundary, which confirms that the
thermal and fossil electrons are well separated in momentum space.

The IC bump extends towards higher momenta for longer compression duration.
In Fig.~\ref{fig:spectrum-convolve}(c) we show $f_\mt{revert}$ computed for
three evenly-spaced times $t=0.47,0.94,1.41 q^{-1}$ in our fiducial simulation.
The spectrum at $t=0.47 q^{-1}$ shows a very weak IC bump, which we attribute
to weaker IC scattering at early times when IC waves are not yet saturated.
The IC bump becomes more prominent at $t=0.94 q^{-1}$ and $1.41 q^{-1}$.
We further explore the link between compression duration and the onset of
scattering at high $p$ later in this manuscript.

We also consider the effect of multiple compress-revert cycles by assuming
that, at the end of each compress-revert cycle, $f_\mt{revert}(p)$ instantly
becomes isotropic in $\mu$; the result is shown in
Fig.~\ref{fig:spectrum-convolve}(d).
Multiple cycles strengthen the IC energy gain betweeen $p=10$ to $100 \;\me c$.
The IC pumping does not extend to $p \gg 100 \;\me c$; CRe with
$p \sim 10^3 \;\me c$ stay adiabatic through multiple compress-revert cycles.
The assumption of instant isotropization between each compress-revert cycle is
questionable; we know from Fig.~\ref{fig:spectrum}(j-l) that the revert spectra
are far from isotropic.
The effect of scattering during decompression, which should bring electrons
closer to isotropy, is left for future work.

\section{Cumulative energy gain from magnetic pumping} \label{sec:energygain}

\begin{figure}
    \includegraphics[width=3.375in]{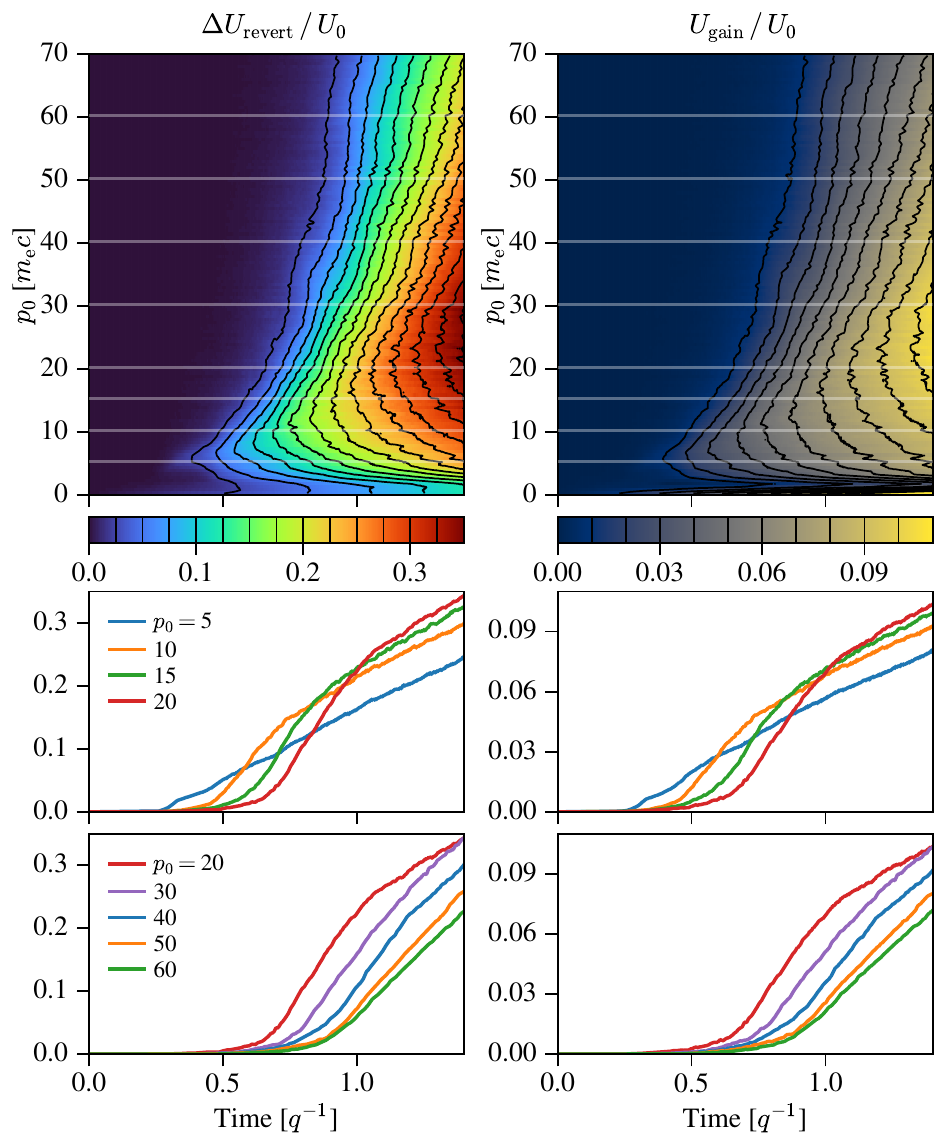}
    \caption{
        Energy gains $\Delta U_\mt{revert} / U_0$ (left column) and
        $U_\mt{gain}/U_0$ (right column), normalized to initial energy.
        Top row: energy gain as a function of CRe initial momentum $p_0$ and
        simulation time $t$, averaged over pitch angle.
        Middle and bottom rows: 1D slices of energy gain as a function of time
        for varying $p_0$.
        Horizontal white lines in top row correspond to $p_0$ selections in
        middle and bottom rows.
    }
    \label{fig:energygain}
\end{figure}

Let us now focus on the efficiency metric $\Delta U_\mt{revert}$, abstracting
away details of the underlying $\mu$-dependent particle spectra.
Fig.~\ref{fig:energygain} shows $\Delta U_\mt{revert}/U_0$ computed for all
test-particle CRe in our simulation, binned by initial CRe momentum with bin
size $\Delta p_0 = 0.5 \;\me c$.
We emphasize three main features.
The lowest-energy CRe, $p_0 \sim 1$--$10 \;\me c$, gain little energy from
magnetic pumping.
Medium-energy CRe, $p_0 \sim 10$--$30 \;\me c$, pump the most efficiently by
virtue of their having initial momenta at or above the expected resonant
$p_{\prll} \sim 4$--$25 \;\me c$ (Eq.~\eqref{eq:scaling}).
The highest-energy CRe, $p_0 \gtrsim 30 \;\me c$, gain energy at later times;
as compression proceeds, CRe of progressively higher $p_0$ ``turn on'' their
energy gain.

We also introduce
$U_\mt{gain}$ to represent the time-integrated energy gain from
all mechanisms other than adiabatic compression, particularly momentum
diffusion.
To compute $U_\mt{gain}$, we decompose each particle's energy gain over a
timestep $\Delta t$ into adiabatic and non-adiabatic
pieces:
\[
    \gamma(t+\Delta t) - \gamma(t) = \Delta\gamma_\mt{gain} + \Delta\gamma_\mt{CGL} \, ,
\]
where $\gamma$ is the particle's Lorentz factor,
\[
    \Delta\gamma_\mt{CGL}
    = \sqrt{1 + {p_\prll(t)}^2 + {p_\perp(t)}^2 \left[B(t+\Delta t)/B(t)\right] }
    - \gamma(t) \, ,
\]
and the remaining energy gain is $\Delta\gamma_\mt{gain}$.
Then we may time integrate and ensemble average to define
\[
    U_\mt{gain}(t) = \left\langle
    \sum_{j=0}^{\lfloor t/\Delta t\rfloor}  \Delta\gamma_\mt{gain}(j\Delta t, \Delta t)
    \right\rangle
\]
shown as a function of $p_0$ and $t$ in Fig.~\ref{fig:energygain}.
The timestep $\Delta t = 4.7 {\Omcio}^{-1}$
matches that used to measure particle scattering in
Sec.~\ref{sec:scatt}.

In Fig.~\ref{fig:energygain} we draw three conclusions concerning
$U_\mt{gain}$.
First, both $U_\mt{gain}$ and $\Delta U_\mt{revert}$ show the same qualitative
features in $(t,p_0)$ coordinates.
We attribute this to the shared gyroresonant nature of both energy gain
processes: $D_{pp}$ for non-adiabatic diffusive energization $U_\mt{gain}$,
and $D_{\mu\mu}$ for magnetic pumping $\Delta U_\mt{revert}$.
Second, the magnitude of $U_\mt{gain}$ is $\abt 10\%$ that of the initial
particle energy by the end of the simulation; however, $U_\mt{gain}(t)$ is
small compared to the total particle energy $U(t)$ arising from compression,
which is $\gtrsim 200\%$ of the initial particle energy $U_0$ by the end of
the simulation.
Finally, $U_\mt{gain}$ decreases as $\vAo/c$ is lowered towards a more
realistic value, whereas $\Delta U_\mt{revert}$ does not vary as strongly with
$\vAo/c$; we show this decrease in $U_\mt{gain}$ later in the manuscript
(Fig.~\ref{fig:dpp-dmumu-scaling-sel}).
On the basis of these observations, we view $U_\mt{gain}$ and hence $D_{pp}$ as
a minor player in CRe energization through our compressive cycle.

\section{Continuous compression controls the efficiency of magnetic pumping} \label{sec:fpmodel}

The 2D structure of $\Delta U_\mt{revert}\left( t, p_0 \right)$ encodes
information about which particles scatter and when they scatter; i.e., it
encodes the time- and $k$-dependent wave spectrum $W(k,t)$, but we lack a
mapping from $W(k,t)$ and $B_g(t)$ to
$\Delta U_\mt{revert}\left(t, p_0\right)$.
To understand the 2D structure of $\Delta U_\mt{revert}$, we perform
Fokker-Planck (F-P) simulations of compression with time-dependent pitch-angle
scattering:
\[
    \frac{\ptl f}{\ptl t}
    + \frac{\dot{B}}{B} \frac{p_\perp}{2} \frac{\ptl f}{\ptl p_\perp}
    = \frac{\ptl}{\ptl \mu} \left(
        D_{\mu\mu} \frac{\ptl f}{\ptl \mu}
    \right)
\]
We sample $280,000$ CRe with momenta between $p_0$ between $0.25$ to
$69.75 \;\me c$ and an isotropic pitch angle distribution
(i.e., uniform $\mu \in [-1,+1]$).
Then, we subject the CRe to the same continuous compression as in our fiducial
simulation, $B_g(t) = B_0 (1+qt)^2$ with $q^{-1} = 800 {\Omcio}^{-1}$, using a
finite-difference method.
Advancing from time $t_n$ to $t_{n+1} = t_n + \Delta t$, each particle's
perpendicular momentum is increased adiabatically as
$p_\perp(t_{n+1}) = p_\perp(t_n) \sqrt{ B_g(t_{n+1})/B_g(t_{n}) }$; the
parallel momentum $p_\prll(t_{n+1}) = p_\prll(t_n)$ is held constant.
The finite-difference timestep $\Delta t = 0.94 {\Omcio}^{-1}$.

At first, the compression is adiabatic to mimic the relatively weak wave power
at early times in our fiducial simulation (Fig.~\ref{fig:overview}(a-c)).
After $t=0.3 q^{-1}$, we begin scattering all particles that satisfy:
\begin{equation} \label{eq:fp-kmin}
    \left| k_\mt{res} \right| = \frac{e B_g(t)}{|\mu| p c} > k_\mt{min}(t)
\end{equation}
where $k_\mt{min}(t)$ is a user-chosen function.
The scattering is implemented as a 1D random walk in pitch angle $\alpha$.
For each time $t_n$, each particle satisfying Eq.~\eqref{eq:fp-kmin} takes a
randomly-signed step $\Delta \alpha = \pm 0.04$ prior to the compression step
$p_\perp(t_n) \to p_\perp(t_{n+1})$.
The variance of the total displacement after $N$ steps is
$\langle \Delta\alpha\Delta\alpha \rangle_\mt{N} = N (\Delta\alpha)^2$,
so the effective diffusion coefficient
$D_{\mu\mu} \sim (1 - \mu^2)
    \langle \Delta\alpha\Delta\alpha \rangle/(2\Delta t)
\approx 8.5 \times 10^{-4} (1-\mu^2) \,\Omcio$.
This $D_{\mu\mu}$ value is weaker than the scattering rate measured in our
fiducial simulation (Fig.~\ref{fig:scattering}(j-l)); nevertheless, the F-P
model returns a comparable value of $\Delta U_\mt{revert}$.
Also, our F-P model deviates from quasi-linear theory in having no $90^\circ$
barrier; particles with $\mu=0$ scatter efficiently in order to mimic the
presence of scattering at $\mu=0$ in Fig.~\ref{fig:scattering}(j-l).
Varying the start time of scattering to either $t=0.0q^{-1}$ or $0.6q^{-1}$
has only a small effect on the F-P model energy gain; the time evolution of
$k_\mt{min}(t)$ is more important.

\begin{figure}
    \includegraphics[width=3.375in]{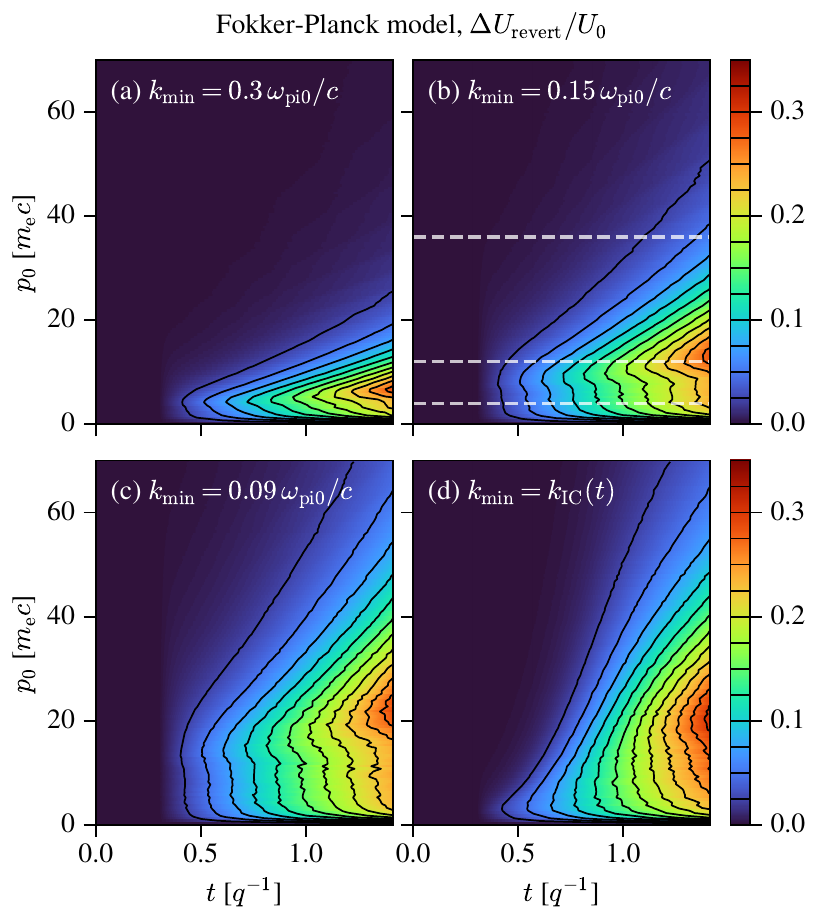}
    \caption{
        Energy gain from a pure pitch-angle scattering Fokker-Planck (F-P)
        model, with different prescriptions $k_\mt{min}$ for the particle
        scattering.
        Particles compress with same $B_g(t)$ as the fiducial PIC simulation.
        For $t < 0.3 q^{-1}$, no particles scatter; compression is adiabatic.
        For $t \ge 0.3 q^{-1}$, particles satisfying Eq.~\eqref{eq:fp-kmin}
        scatter by random walk in pitch-angle; see text for details.
        (a-c): Time-constant $k_\mt{min} = 0.3$, $0.15$, and $0.09\;\ompio/c$.
        (d): Time-dependent $k_\mt{min} = k_\mt{IC}(t)$
        (Eq.~\eqref{eq:kicdrift}) mimicking the decreasing-$k$ drift of ion
        cyclotron wave power seen in our fiducial PIC simulation.
        Dashed white lines in (b) correspond to particle samples in
        Fig.~\ref{fig:fpmodel-explainer}.
    }
    \label{fig:fpmodel}
\end{figure}

\begin{figure}
    \centering
    \includegraphics[width=2.5in]{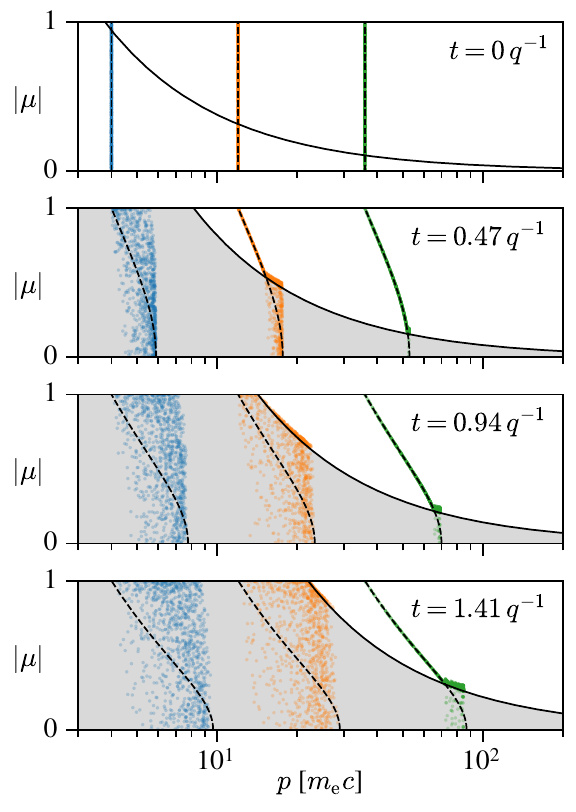}
    \caption{
        Evolution of initially isotropic, monoenergetic particle distributions
        with $p_0 = 4, 12, 36$ in F-P scattering model with
        $k_\mt{min} = 0.15 \;\ompio/c$, corresponding to
        Fig.~\ref{fig:fpmodel}(b).
        Black solid curves show Eq.~\eqref{eq:fp-kmin} bounds; particles
        scatter within the bounded, gray-shaded region and evolve adiabatically
        otherwise.
        Black dashed lines mark curves upon which particles would evolve from
        $p_0$ if there were no scattering at all.
    }
    \label{fig:fpmodel-explainer}
\end{figure}

\begin{figure*}
    \includegraphics[width=7.1in]{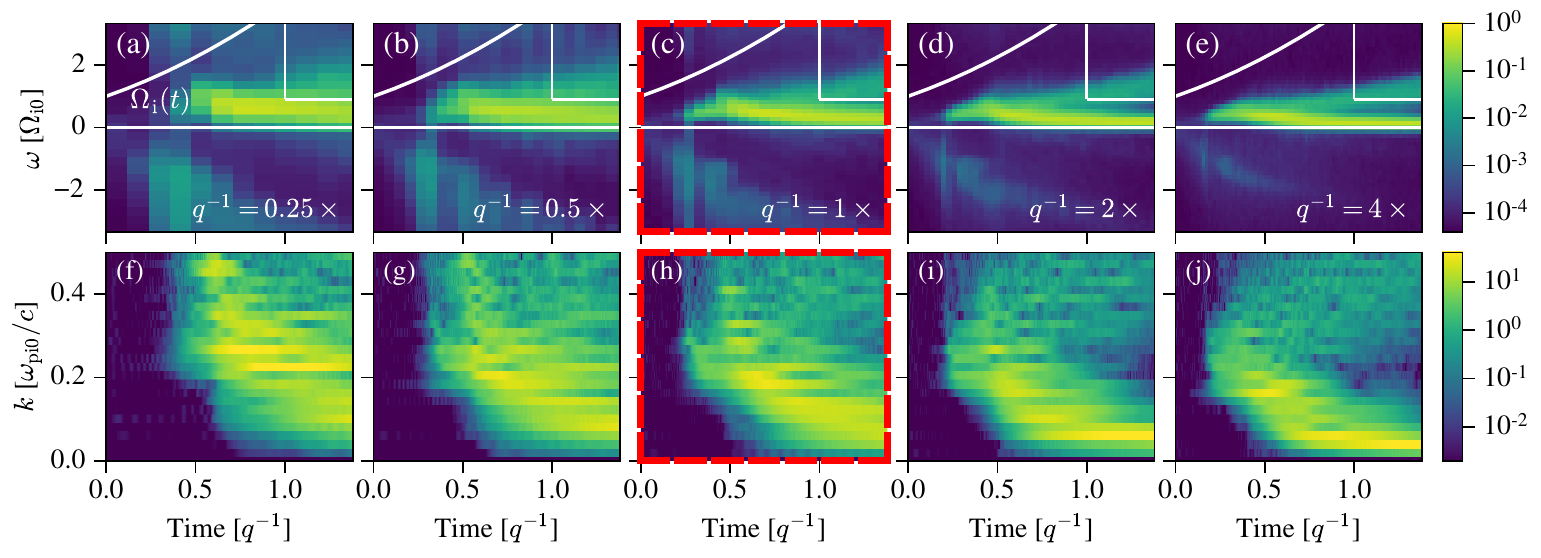}
    \caption{
        Wave power spectrogram of $B_z + i B_x$ for varying $q$,
        from small $q^{-1}$ / fast compression (left)
        to large $q^{-1}$ / slow compression (right).
        (a-e) Spectrogram in $(t,\omega)$ coordinates with a finite
        time-binning for each pixel.
        White curve in top left of each panel is $\Omci(t)$.
        Within the white boxes ($t > 1 q^{-1}$ and $\omega/\Omcio > 0.9$), we
        average the power spectral density (PSD) to estimate the power due to
        unstable IC waves at high $\omega$, omitting the linearly-stable IC
        waves at low $\omega$.
        (f-j) Wave power spectrum in $(t,k)$ coordinates without time binning.
        Red dashed frames (c,h) mark fiducial simulation; i.e., same data as
        Fig.~\ref{fig:overview}.
    }
    \label{fig:db-qinv-fft}
\end{figure*}

Figure~\ref{fig:fpmodel} shows the magnetic-pumping energy gain in our F-P
model for four different choices of $k_\mt{min}$.
We first consider constant $k_\mt{min} = 0.3$, $0.15$, and $0.09 \;\ompio/c$ in
Fig.~\ref{fig:fpmodel}(a-c).
Then, we adopt a time-dependent $k_\mt{min} = k_\mt{IC}(t)$, using
Eq.~\eqref{eq:kicdrift} to mimic the decreasing-$k$ drift of ion cyclotron wave
power in our fiducial PIC simulation.
We draw three conclusions.
First, the magnetic-pumping energy gain has a self-similar geometric structure
in $(t,p_0)$ coordinates for $k_\mt{min}$ constant in time; changing
$k_\mt{min}$ is the same as rescaling $p_0$ by a factor $1/k_\mt{min}$
(Eq.~\eqref{eq:fp-kmin}), so the panels of Fig.~\ref{fig:fpmodel}(a-c) are
identical up to linear rescaling along the $y$-axis.
Second, the particles gaining the most energy from magnetic pumping have $p_0$
somewhat higher than the initial resonant $p_\prll$ at $t=0$.
For example, choosing $k_\mt{min}=0.09\ompio/c$ gives the most energy to
particles with $p_0 \approx 20$--$30 \;\me c$ (Fig.~\ref{fig:fpmodel}(c)),
whereas Eq.~\eqref{eq:fp-kmin} requires $p_{\prll} \le 6$--$36 \;\me c$.
Third, the time-dependent $k_\mt{min} = k_\mt{IC}(t)$ broadens the energy-gain
``resonance'' feature in $\Delta U_\mt{revert}$ towards higher $p_0$
(Fig.~\ref{fig:fpmodel}(d)).

To understand how magnetic pumping interacts with continuously-driven
compression to ``select'' a range of $p_0$ with the highest magnetic pumping
efficiency, Fig.~\ref{fig:fpmodel-explainer} shows how isotropic, monoenergetic
particle distributions with $p_0 = 4, 12, 36\;\me c$ evolve over time while
subjected to both compression and pitch-angle scattering (after $t = 0.3
q^{-1}$) for all particles with $k_\mt{min} = 0.15 \ompio/c$
(Fig.~\ref{fig:fpmodel}(b)).
The lowest-energy particles, $p_0=4\;\me c$ (blue), scatter promptly at all
pitch angles from $t \ge 0.3 q^{-1}$ and onwards, so the magnetic pumping is
less efficient.
The medium-energy particles, $p_0=12\;\me c$ (orange), only scatter near
$\mu=0$ at early times $t\sim 0.3q^{-1}$, but their scattering extends to most
$\mu$ values by the simulation's end.
The highest-energy particles $p_0=36\;\me c$ (green) are mostly adiabatic; few
such particles scatter until later times, so their energy gain from magnetic
pumping is small.

Preferential scattering near $\mu=0$, where compression gives the most energy
(as compared to larger $|\mu|$), causes the medium-energy particles to migrate
to large $|\mu|$ and ``lock in'' their compressive energy gain; therefore,
medium-energy particles participate most efficiently in magnetic pumping.
We interpret orange particles accumulating at the scattering region boundaries
in Fig.~\ref{fig:fpmodel-explainer}, as well as the skewed particles at large
$|\mu|$ in Fig.~\ref{fig:spectrum}(j-l), as evidence for energy locking.
The highest-energy particles also scatter from $\mu \sim 0$ towards the
scattering boundary (Fig.~\ref{fig:fpmodel-explainer}), but (1) fewer particles
are able to participate, and (2) the smaller $\mu$ of the scattering boundary
causes more compressive energy gain to be removed in decompression.
The lowest-energy particles, because they scatter at all $\mu$, easily flow
between $\mu \sim 0$ and $|\mu| \sim 1$; there is no region of $(p,\mu)$ space
in which particles may lock energy gained from $\mu \sim 0$.

The drift of IC power towards low $k$ further modifies particle energization.
In Fig.~\ref{fig:fpmodel-explainer}, the gray scattering region expands
rightwards as time progresses: $p_\parallel \propto B_g(t) / k_\mt{res}$, and
$k_\mt{res}$ decreasing in time will hasten that expansion and therefore widen
the band of medium-energy particles.
Previously, \citet[Sec.~4]{matsukiyo2009} have also noted how Alfv\'{e}nic
waves drifting to low $k$ may help accelerate particles that can stay within
the range of resonant momenta of the time-evolving waves.

Compression and the drift of IC power towards low $k$ together can thus
explain, qualitatively, the distinct low-, medium-, and high-energy CRe
structure of $\Delta U_\mt{revert}$ as a function of $t$ and $p_0$
(Fig.~\ref{fig:energygain}).

\section{Compression Rate Dependence} \label{sec:qinvscaling}

\begin{figure}
    \includegraphics[width=3.375in]{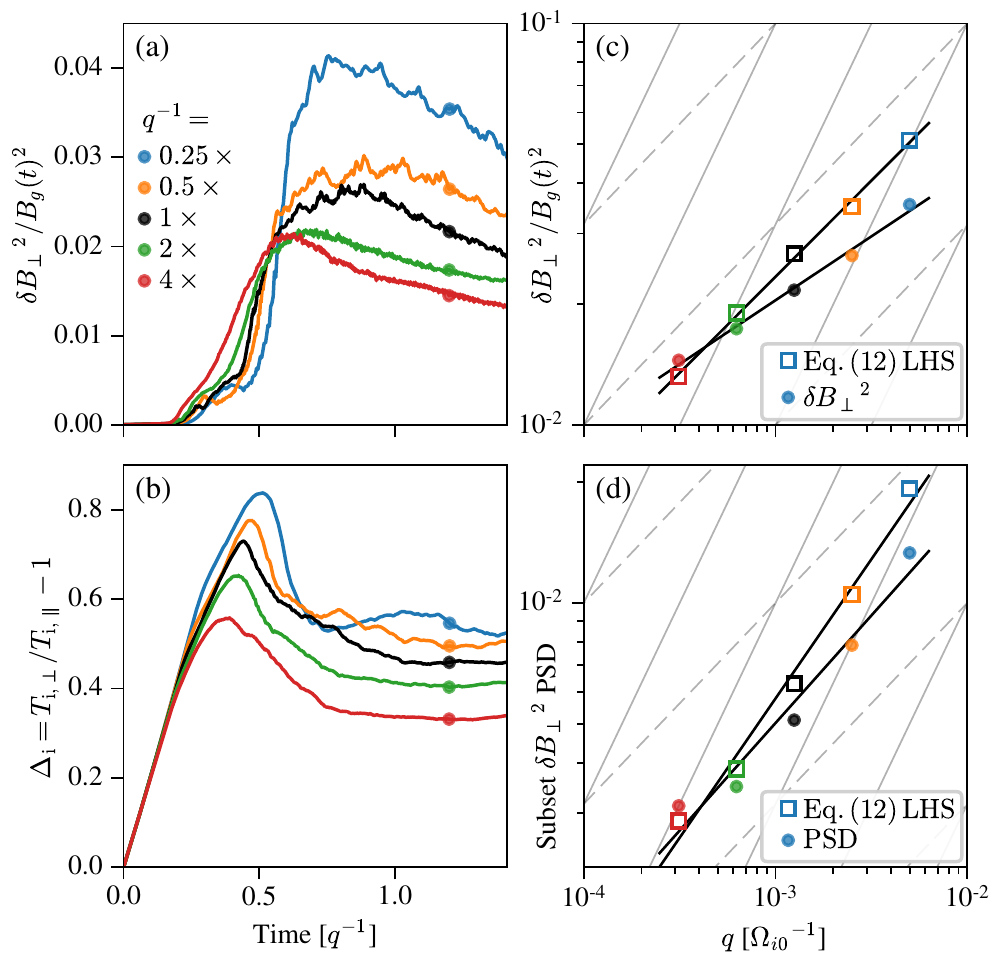}
    \caption{
        Scaling of $\delta {B_\perp}^2/B_g(t)^2$ power from simulations with
        varying $q$, indicated by marker/line color.
        (a) Wave power $\delta {B_\perp}^2/B_g(t)^2$ over time, domain-averaged.
        (b) Ion temperature anisotropy
        $\Delta_\mt{i} = T_\mt{i\perp}/T_\mt{i\prll}-1$ over time $t$.
        (c) Wave power at $t = 1.2\,q^{-1}$ plotted as a function of $q$
        (solid circles).
        And, the same wave power multiplied by
        $\Delta_\mt{i} (2\Delta_\mt{i}+3))(\Delta_\mt{i}+1)$, i.e., the
        left-hand side (LHS) of Eq.~\eqref{eq:driftkinetic3} (hollow squares),
        to test the linear $q$ scaling of Eq.~\eqref{eq:driftkinetic3}.
        Solid, dashed light-gray lines are $\propto q$, $\sqrt{q}$ scalings
        respectively.
        Solid black lines are least-squares power-law fits.
        (d) Like (c), but replace $\delta {B_\perp}^2/B_g(t)^2$ with
        time-averaged power spectral density (PSD) sampled from white-boxed
        subsets of the spectrograms in Fig.~\ref{fig:db-qinv-fft}(a-e); see
        text for details.
    }
    \label{fig:db-qinv}
\end{figure}

\begin{figure*}
    \includegraphics[width=7.1in]{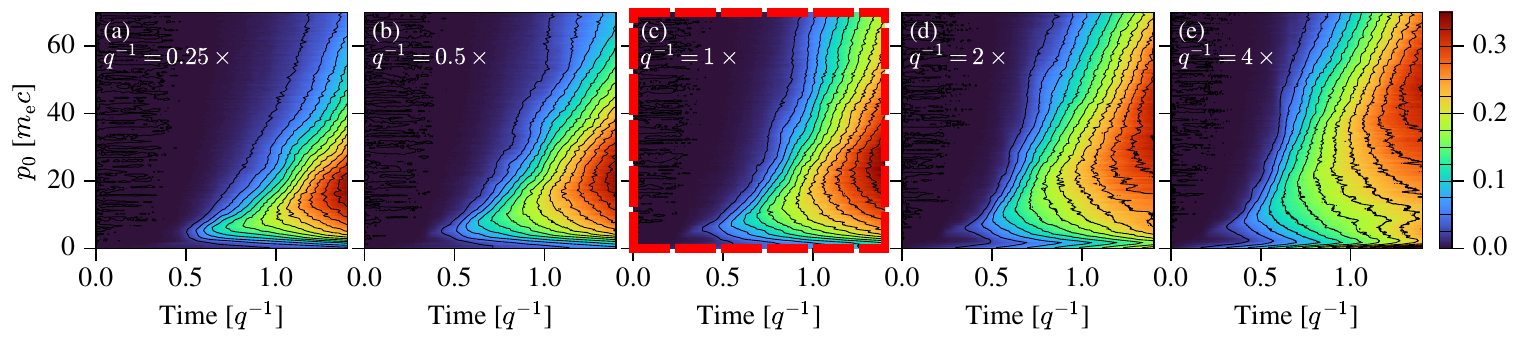}
    \caption{
        Energy gain metric $\Delta U_\mt{revert}/U_0$, showing effect of
        (a) faster $q^{-1} = 200\,\Omcio^{-1}$ to
        (e) slower $q^{-1} = 3200\,\Omcio^{-1}$
        compression upon CRe energization.
        Red dashed panel is fiducial simulation, same as
        Fig.~\ref{fig:energygain}.
    }
    \label{fig:energygain-qinv}
\end{figure*}

In our simulations, the compression timescale $q^{-1} = 800\,{\Omcio}^{-1}$
corresponds to $q^{-1} \sim 10^{-3} \unit{year}$ if one assumes
$B_0 = 3 \unit{\mu G}$, which is much smaller than the actual sound-crossing
time $\abt 10^8 \unit{year}$ for cluster-scale ICM bulk motion.
How do the CRe energy gain and the IC wave spectrum change with $q^{-1}$ in our
simulations?
For larger $q^{-1}$, linearly-unstable IC waves grow earlier and attain smaller
$k$ at late times (Fig.~\ref{fig:db-qinv-fft}), so we expect the IC wave
resonance to broaden towards higher $p$.

We also expect the wave power $\delta B_\perp^2$ to weaken for larger $q^{-1}$
per Eq.~\eqref{eq:driftkinetic2}, which may be rewritten more explicitly as
\begin{equation} \label{eq:driftkinetic3}
    \left(\frac{\delta B_\perp}{B_g}\right)^2
    \frac{\Delta_\mt{i} (2\Delta_\mt{i}+3)}{\Delta_\mt{i}+1}
    \propto
    \frac{2q/\Omcio}{(1+qt)^3} \, .
\end{equation}

In Fig.~\ref{fig:db-qinv}, we check if the linear scaling with $q$ predicted by
Eq.~\eqref{eq:driftkinetic3} holds in our simulations.
Both $(\delta B_\perp/B_g)^2$ and $\Delta_\mt{i}$ decrease when $q$ decreases
(Fig.~\ref{fig:db-qinv}(a-b)).
At $t=1.2q^{-1}$, we sample and plot $(\delta B_\perp/B_g)^2$ as a function of
$q$ (Fig.~\ref{fig:db-qinv}(c), solid markers).
We similarly compute and plot the left-hand side (LHS) of
Eq.~\eqref{eq:driftkinetic3} (Fig.~\ref{fig:db-qinv}(c), hollow markers).
Both quantities appear to follow a power law scaling $q^n$ with exponent
$n \lesssim 0.5$, which is a weaker proportionality than predicted by
Eq.~\eqref{eq:driftkinetic3}.

Waves at differing $k$ may not contribute equally towards balancing
compression-driven anisotropy;
recall how the strongest waves lie outside the unstable $\omega$ range in
Fig.~\ref{fig:overview}(a), and how Eq.~\eqref{eq:driftkinetic2} agrees better
with the unstable wave power rather than the total wave power in
Fig.~\ref{fig:overview}(c).
We thus suspect that low-frequency wave power may participate less in
regulating the ion anisotropy.
Does the anisotropy-driven high-frequency wave power, rather than total wave
power, scale linearly with $q$ per Eq.~\eqref{eq:driftkinetic3}?
We select wave power with $\omega/\Omcio > 0.9$ by computing the average wave
power spectral density (PSD) in the top-right white boxes of
Fig.~\ref{fig:db-qinv-fft}(a-e) panels;\footnote{
    The PSD averaged in Fourier space equals the real-space average of $(\delta
    B_{\perp}/B_g)^2$ (i.e., an $\omega$-average of
    Fig.~\ref{fig:db-qinv-fft}(a-e) or a $k$-average of
    Fig.~\ref{fig:db-qinv-fft}(f-j) will return the domain-averaged wave power
    in Fig.~\ref{fig:db-qinv}(a)).
}
the resulting PSD is plotted against $q$ in Fig.~\ref{fig:db-qinv}(d).
The PSD multiplied by $\Delta_\mt{i} (2\Delta_\mt{i}+3)/(\Delta_\mt{i}+1)$
appears to follow a power-law scaling $q^n$ with exponent $n$ between
0.5 and 1.

Least-squares fits of form $A \left(q/\Omcio\right)^n$, with free parameters
$A$ and $n$, are plotted as solid black lines in Fig.~\ref{fig:db-qinv}(c-d).
For Eq.~\eqref{eq:driftkinetic3} LHS
(Fig.~\ref{fig:db-qinv}(c), hollow squares),
and $\delta {B_\perp}^2$
(Fig.~\ref{fig:db-qinv}(c), solid circles),
we obtain $n = 0.48 \pm 0.02$ and $0.32 \pm 0.02$ respectively.
For the high-frequency wave $\mt{PSD}$ replacing $\delta {B_\perp}^2$ in
Eq.~\eqref{eq:driftkinetic3} LHS
(Fig.~\ref{fig:db-qinv}(d), hollow squares),
and the high-frequency wave $\mt{PSD}$ alone
(Fig.~\ref{fig:db-qinv}(d), solid circles),
we obtain $n = 0.69 \pm 0.05$ and $0.54 \pm 0.07$ respectively.
We fit the data in log coordinates (i.e., linear regression).
The uncertainty on $n$ is one standard deviation estimated by assuming
$\chi^2_\mt{reduced} = 1$, as no data uncertainty is used in fitting.
We expect that the systematic uncertainty is larger.

We warn that our $\omega/\Omcio > 0.9$ threshold does not cleanly separate low-
and high-frequency wave power for every simulation because the $\omega$ range
of the wave power varies with $q$ (Fig.~\ref{fig:db-qinv-fft}).
Altering the $\omega$ threshold will also alter the $q$-scaling exponent in
Fig.~\ref{fig:db-qinv}(d).
A multi-component fit to the power spectrum may better separate the low- and
high-frequency wave power and so provide a better test of
Eq.~\eqref{eq:driftkinetic2}, but we omit such detailed modeling for now.

We also show how the CRe energy gain $\Delta U_\mt{revert}$ changes with $q$ in
Fig.~\ref{fig:energygain-qinv}.
As $q^{-1}$ increases, the optimal $p_0$ range for magnetic pumping both widens
and moves to higher momenta, which we ascribe to both the lower late-time $k$
and earlier onset of waves with respect to compression timescale $q^{-1}$.
We suspect that wave evolution towards lower $k$ is the dominant effect
altering the shape of $\Delta U_\mt{revert}$ for varying $q^{-1}$.
We do not observe, by eye, a trend in the peak magnitude of $\Delta
U_\mt{revert}$ with respect to $q$.

\section{Scaling to realistic ICM plasma parameters} \label{sec:vAc-mime}

\begin{figure*}
    \includegraphics[width=7.1in]{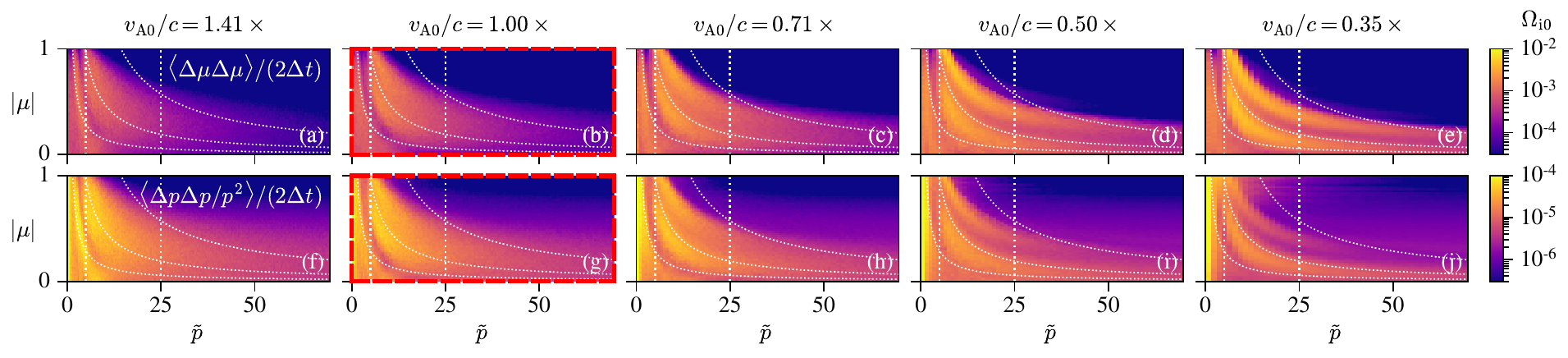}
    \caption{
        Scattering rates measured at $t=0.35 q^{-1}$ in simulations of varying
        $\vAo/c$, decreasing left to right.
        (a-e) Pitch-angle scattering
        $\langle \Delta \mu \Delta \mu \rangle/(2\Delta t)$.
        (f-j) Momentum scattering
        $\langle \Delta p \Delta p / p^2 \rangle/(2\Delta t)$.
        White dotted vertical lines mark averaging region used in
        Fig.~\ref{fig:dpp-dmumu-scaling}.
        White dotted curves are same contours of constant resonant wavenumber
        as in Fig.~\ref{fig:scattering}.
        Red dashed frames around (b,f) mark fiducial simulation.
        All rates in units of $\Omcio$.
    }
    \label{fig:dpp-dmumu-scaling-sel}
\end{figure*}

\begin{figure*}
    \includegraphics[width=7.1in]{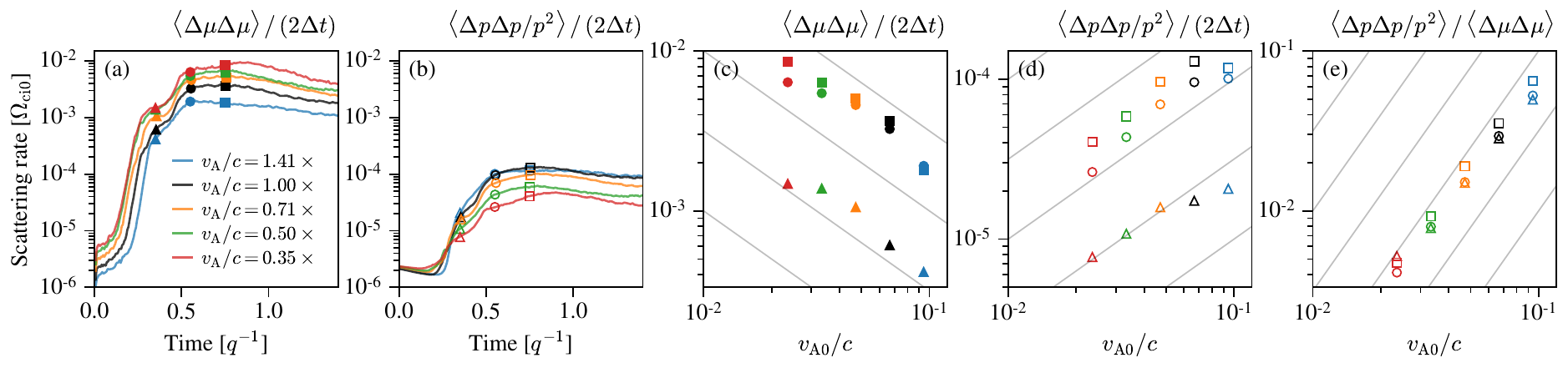}
    \caption{
        (a-b) Scattering rates
        $\langle \Delta \mu \Delta \mu \rangle/(2\Delta t)$ and
        $\langle \Delta p \Delta p / p^2 \rangle/(2\Delta t)$ measured in five
        simulations with varying $\vAo/c$ (line/marker color), reported in
        units of $\Omcio$.
        Rates are averages over $(|\mu|,\tilde{p})$ regions marked in
        Fig.~\ref{fig:dpp-dmumu-scaling-sel}.
        (c-d) Sample points from (a-b) plotted as a function of $\vAo/c$.
        Light gray lines show $(\vA/c)^{-1}$ (panel c) and $(\vA/c)^{+1}$
        (panel d) scalings.
        (e) Ratio of $p$ and $\mu$ scattering rates.
        Light gray lines show $(\vA/c)^2$ (panel e) scaling.
        In all panels, symbols correspond to different times at which
        scattering rates are measured.
    }
    \label{fig:dpp-dmumu-scaling}
\end{figure*}

\begin{figure*}
    \centering
    \includegraphics[width=6in]{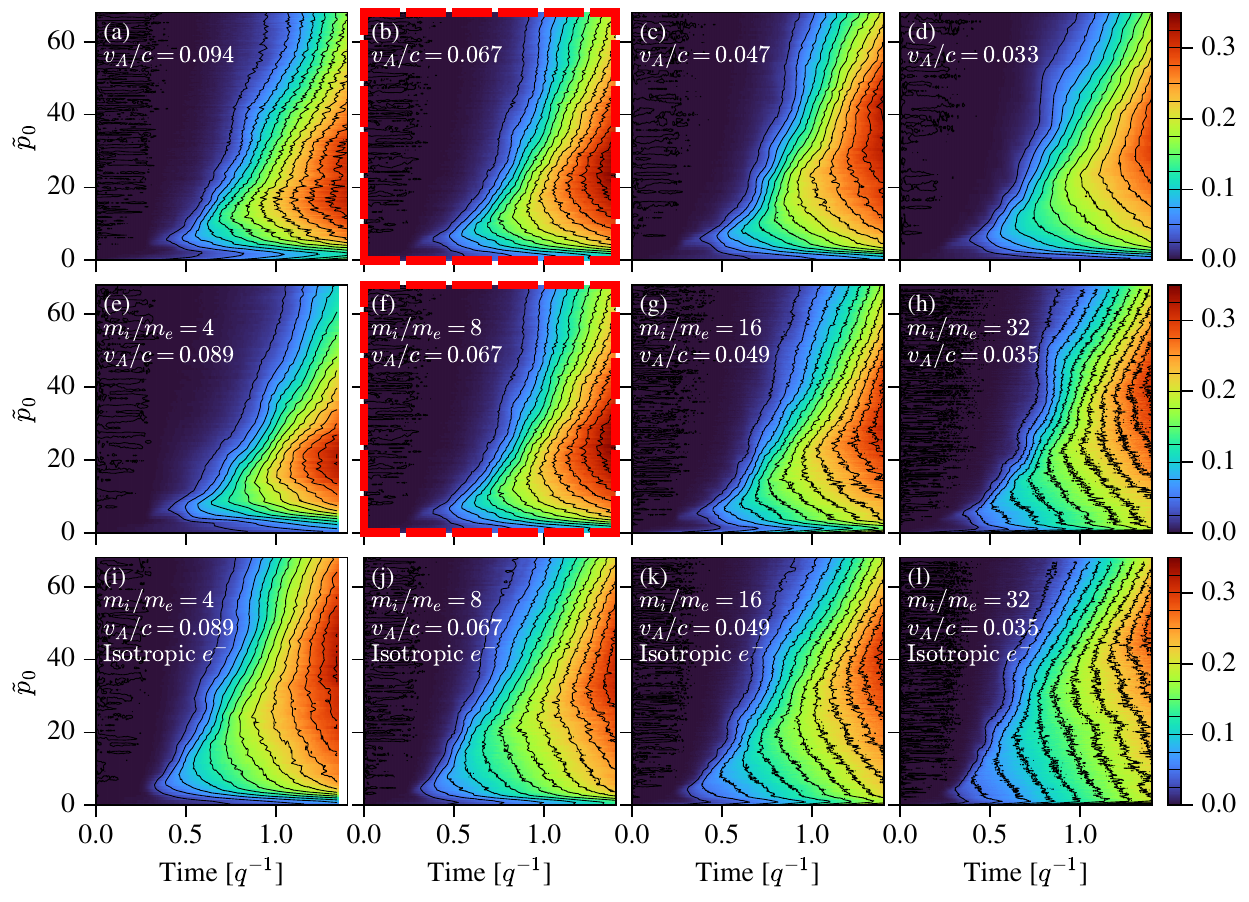}
    \caption{
        Energy gain metric $\Delta U_\mt{revert}/U_0$, like
        Fig.~\ref{fig:energygain} but with $\tilde{p}_0$ rather than $p_0$ on
        the y-axis, for simulations of varying $\vAo/c$ and $\mime$.
        (a-d): Vary $\vAo/c$ with fixed $\mime=8$.
        (e-h): Vary $\mime$ and $\vA/c$ together, to help understand the effect
        of $\mime$.
        (i-l): Like (e-h), but disable whistler waves by compressing electrons
        isotropically at the rate $q_\mathrm{iso}=2q/3$.
        Red-dash framed panels (b,f) are the fiducial simulation, previously
        shown in Fig.~\ref{fig:energygain}.
    }
    \label{fig:energygain-vAc-mime}
\end{figure*}

How do more realistic simulation parameters (higher $\mi/\me$, lower $\vA/c$)
alter our results?
Let us define a dimensionless CRe momentum
\[
    \tilde{p} \propto p \left(\frac{\mi}{\me}\right)^{-1}
                        \left(\frac{\vAo}{c}\right)^{-1}
\]
with the constraint $\tilde{p} = p$ for our fiducial simulation parameters,
motivated by the gyro-resonance scaling (Eq.~\eqref{eq:scaling});
recall that $v_\mt{th,i}/c \propto \vA/c$ for fixed $\beta_\mt{p}$.
As in Secs.~\ref{sec:spectrum}--\ref{sec:energygain}, $\tilde{p}_0$ is the
value of $\tilde{p}$ for CRe particles at $t=0$.
If simulations of varying $\vAo/c$ and $\mime$ have a similar IC wave spectrum
$W(t,k)$ for fixed $q/\Omcio$, then particle scattering and energization should
also have a similar structure in $\tilde{p}$.

We vary $\vAo/c$ of our fiducial simulation by factors of $\sqrt{2}$ and
measure particle scattering rates
$\langle\Delta p \Delta p/p^2\rangle/(2\Delta t)$ and
$\langle\Delta\mu\Delta\mu\rangle/(2\Delta t)$ in discrete $(\mu,p)$ bins.
As in Sec.~\ref{sec:scatt}, the timestep $\Delta t \approx 5 {\Omcio}^{-1}$.
The momentum bin width $0.5 \me c$ is fixed for all simulations, so the plotted
$\tilde{p}$ bin width varies between simulations in
Fig.~\ref{fig:dpp-dmumu-scaling-sel}.

The measured scattering rates indeed have similar shape in $(|\mu|,\tilde{p})$
coordinates for varying $\vAo/c$ (Fig.~\ref{fig:dpp-dmumu-scaling-sel}).
At lower $\vAo/c$, a double-lobed scattering region appears along the resonant
contours.
Lower $\vAo/c$ also alters the apparent edge of the scattering region at
$\tilde{p} > 25$ towards possibly better agreement with the predicted resonant
contours from Eq.~\eqref{eq:qlt-kres}, although the scattering region edge
still disagrees at low $\tilde{p} < 25$.

To explore how scattering scales with $\vAo/c$, we average scattering rates
over $|\mu|$ and $\tilde{p} \in [5,25]\me c$ to sample the strongest IC wave
signal in momentum space.
The average rates are plotted as a function of time in
Fig.~\ref{fig:dpp-dmumu-scaling}(a-b); the same rates sampled at three discrete
times are then plotted as a function of $\vAo/c$ in
Fig.~\ref{fig:dpp-dmumu-scaling}(c-d).
The pitch-angle and momentum scattering rates increase and decrease,
respectively, as $\vAo/c$ decreases.
We interpret the data as showing a transition from mildly relativistic to
non-relativistic behavior as we lower $\vAo/c$.
At lower $\vAo/c$ than shown, we expect that the pitch-angle scattering should
become independent of $\vAo/c$, while momentum scattering should scale as
$(\vAo/c)^2$.
We also verify the expected QLT scaling:
\[
    \frac{ \langle \Delta p \Delta p / p^2 \rangle }
    { \langle\Delta\mu\Delta\mu\rangle }
    \propto \left( \frac{\vA}{c} \right)^2
\]
in Fig.~\ref{fig:dpp-dmumu-scaling}(e), which shows a power-law-like scaling
consistent through the entire range of $\vAo/c$ considered.

As previously claimed, momentum scattering is not important in a single
compress-revert cycle for our simulation parameters.
We see that $\langle\Delta p \Delta p/p^2\rangle/(2\Delta t)$ is
$\abt 10^{-2}\times$ smaller than
$\langle\Delta\mu\Delta\mu\rangle/(2\Delta t)$,
and the QLT scaling assures us that momentum scattering is even less important
in real ICM with $\vA/c \lesssim 10^{-3}$.
In Fig.~\ref{fig:dpp-dmumu-scaling}(e), the separation between data measured at
different times in the same simulation may be partly attributed to time
variation in $\vA(t)/c$.

We proceed to vary $\mime$ and $\vAo/c$ together, now focusing solely on the
magnetic pumping efficiency $\Delta U_\mt{revert}$, in
Fig.~\ref{fig:energygain-vAc-mime}.
Across all panels, we observe a similar three-band structure as in our fiducial
simulation:
low-energy CRe ($\tilde{p}_0 \lesssim 5$) gain little energy,
medium-energy CRe ($\tilde{p}_0 = 10$--$30$) gain the most energy,
and high-energy CRe ($\tilde{p}_0 \gtrsim 30$) progressively ``turn on'' their
energy gain over time, later for higher energy CRe.
If we remove whistler waves by compressing electrons isotropically
(Sec.~\ref{sec:methods}),
comparing
Fig.~\ref{fig:energygain-vAc-mime}(e-h) against
Fig.~\ref{fig:energygain-vAc-mime}(i-l):
the region of most efficient energy gain shifts to higher $\tilde{p}_0$,
and the maximum value of $\Delta U_\mt{revert}/U_0$ decreases in magnitude
by $\abt 0.05$.
Otherwise, the overall shape of $\Delta U_\mt{revert}$ remains similar
when comparing simulations with and without whistler waves.

\section{Conclusions and Outlook}

We have used 1D PIC simulations to show how ICM fossil CRe gain energy from
bulk compression by scattering upon IC waves excited by anisotropic thermal
ions.
The energy gain comes from magnetic pumping, and we have measured the
momentum-dependent pumping efficiency.
Some summary points follow.
First, high-$\bp$ plasma microinstabilities have a convenient wavelength --
comparable to the Larmor radius of thermal protons -- to interact with and
scatter fossil CRe in the ICM of galaxy clusters.
Second, continuous compression and wave-power drift towards low $k$ both
increase, over time, the CRe momentum $p$ that can resonantly scatter on IC
waves and hence gain energy via magnetic pumping.
The increase in resonant $p$ may be viewed as a time-delayed scattering for
high-$p$ CRe, which can help increase the pumping energy gain compared to
continuous scattering from beginning to end of the simulation.
Third, IC wave pumping is robust with respect to mass ratio $\mi/\me$ and
$\vAo/c$ and is not sensitive to the presence or absence of whistler waves
driven by thermal electrons.
Although the simulated $\mime$ and $\vAo/c$ are not realistic, the lower
$\mime$ and higher $\vAo/c$ cancel such that simulated resonant momenta are
only $2$--$3\times$ lower than real fossil CRe.

Our 1D setup with an adiabatic ``revert'' is unrealistic in some ways.
The compression factor $\abt 6$ at the end of our simulation exceeds the
expected density contrast of both weak ICM shocks and subsonic compressive ICM
turbulence \citep[e.g.,][]{gaspari2013}.
More realistic, non-adiabatic decompression may excite firehose modes that
should also resonantly scatter CRe and alter $\Delta U_\mt{revert}$
\citep{melville2016,riquelme2018,ley2022}.
In 2D or 3D simulations, the low-$k$ drift of IC wave power may not persist,
and mirror modes may weaken IC waves; both effects will weaken the energy gain
from IC wave pumping.
Nevertheless, magnetic pumping via resonant scattering on firehose fluctuations
or non-resonant scattering on mirror modes remains possible, for both firehose
and mirror modes will also have a convenient wavelength to interact with fossil
CRe.
Varying $|\vec{B}|$ in solenoidal, shear-deforming flows will also excite the
same high-$\bp$ plasma microinstabilities to scatter and magnetically pump CRe.

Our treatment of a collisionless ion-electron plasma has neglected (1)
Coulomb collisions, and (2) the presence of heavier ions.
Regarding (1), the collision rate varies within a cluster.
The ICM density decreases to $\abt 10^{-5}$--$10^{-4} \unit{cm^{-3}}$ at
large radii from cluster centers, and the proton collision time can there
reach $\gtrsim 100$ Megayears, comparable to the sound-crossing time
as discussed in Sec.~\ref{sec:qinvscaling}.
In denser gas closer to cluster centers, collisions may inhibit large-scale
eddies from driving particle anisotropy.
But, we expect that the turbulent cascade will eventually reach an eddy
scale where the turnover rate is faster than the collision rate, so that
particle anisotropy may be collisionlessly driven.
Regarding (2), He and heavier ions are known to exist in the ICM
\citep{abramopoulos1981,peng2009,berlok2015,mernier2018}.
He++ and other ions will modify the parallel plasma dispersion relation
\citep{smith1964} and proton cyclotron instability growth rate
\citep{gary1993}, and He++ cyclotron waves may themselves be excited
\citep{gary1994-he}.
Mirror and firehose linear instability thresholds will be altered as well
\citep{hellinger2007-mirror,chen2016}.
The precise wave spectrum and hence CRe energy gain would thus change, but
we expect that CRe may still gain energy by magnetic pumping in the
presence of heavier ICM ions.

How does CRe energization by high-$\bp$ IC wave magnetic pumping fit into the
broader context of large-scale ICM flows and turbulence?
At ion Larmor scales, we expect power from high-$\bp$ plasma
micro-instabilities to be much larger than power from the direct turbulent
cascade.
Let us suppose that the ICM has a turbulent magnetic energy spectrum:
\[
    \left\langle \frac{B^2}{8\pi} \right\rangle
    = \frac{1}{V} \int \frac{B^2}{8\pi} \dtl V
    = \int_{2\pi/L}^\infty W_\mt{turb}(k) \dtl k
    \propto \int k^{-n} \dtl k
\]
with outer scale $L$ and $n = 5/3$ for a Kolmogorov cascade.
The energy at the ion (proton) Larmor wavenumber $k_\mt{i} = 2\pi/\rLi$ may be
estimated as \citep{kulsrud1969}:
\begin{equation} \label{eq:wturb}
    \frac{
        k_\mt{i} W_\mt{turb}(k_\mt{i})
    }{
        \left\langle B^2/(8\pi) \right\rangle
    }
    = (n-1) \left( \frac{L}{\rLi} \right)^{-n+1}
    \approx 6.7 \times 10^{-11}
\end{equation}
for ICM parameters $L = 1 \unit{Mpc}$ and $\rLi = 1 \unit{npc}$.
For comparison, our fiducial simulation has:
\begin{equation} \label{eq:wpic}
    \frac{
        k_\mt{i} W_\mt{PIC}(k_\mt{i})
    }{
        {B_g}^2/(8\pi)
    }
    \approx 5.1 \times 10^{-3} \, ,
\end{equation}
We suppose that the simulated $B_g^2$ corresponds to the total magnetic energy
$\langle B^2 \rangle$ in the ICM, because energy resides at the largest scales
in the Kolmogorov spectrum.

Let us further consider IC waves driven by a compressive eddy at a galaxy
cluster's outer scale, $\abt 1\unit{Mpc}$.
Using the estimate from the beginning of Sec.~\ref{sec:qinvscaling}, the
compression timescale $q^{-1}$ will be $10^{11}\times$ larger than in our
simulation.
Combined with the scaling $\delta B_\perp^2 \propto q^{0.28}$ from
Fig.~\ref{fig:db-qinv}(c), we should decrease our estimate of
$k_\mt{i} W_\mt{PIC}(k_\mt{i})$ in Eq.~\eqref{eq:wpic} by a factor of $10^3$ in
order to extrapolate to realistic conditions.
The IC wave power so extrapolated remains $3\times10^5$ times larger than the
power expected from the turbulent direct cascade at ion Larmor scales.

The excess power at ion Larmor scales may also contribute to stochastic
re-acceleration via momentum scattering ($D_{pp}$), as explored for
Alfv\'{e}nic cascades by \citet{blasi2000,brunetti2004}.
Let us suppose that $D_{pp} \propto (\vA/c)^2 q^{0.28}$, from
Fig.~\ref{fig:dpp-dmumu-scaling} and its accompanying discussion.
Again, take ICM outer scale $q^{-1} \sim 10^{11}\times$ larger than our
simulation, and also take ICM $\vA/c \sim 10^{-3}$ and
${\Omci}^{-1} \sim 10^{-6} \unit{year}$.
Our measured momentum scattering then extrapolates to
$\langle \Delta p \Delta p / p^2 \rangle/(2\Delta t) \sim 10^{-11} \Omci$.
The corresponding acceleration time $\abt 10^5 \unit{year}$ is short compared
to cosmological timescales.

What is the efficiency of magnetic pumping, as well as stochastic
re-acceleration, upon IC waves in this slowly-forced, turbulent setting?
A quantitative answer is beyond the scope of this work, but we make a few
remarks.
For CRe momenta within the band of IC wave resonance, scattering will occur
quickly and persist throughout the bulk compression.
Both resonant magnetic pumping and stochastic re-acceleration will be limited
by the available IC wave bandwidth, so electrons will not reach arbitrarily
high energies.
If the IC wave drift rate towards low $k$ scales with $\Omci$ rather than $q$,
owing to the smaller $q$ in reality, wave energy may continue cascading to
smaller $k$ than in our simulations and so help scatter and pump CRe at even
higher momenta.
At galaxy cluster merger shocks, non-thermal protons may also alter the growth
and damping of IC waves and hence their resulting bandwidth
\citep[e.g.,][]{dos-santos2015}.
In a turbulent flow, the microinstabilities will not be volume filling; CRe
streaming in and out of the scattering regions may also alter the energy gain
from magnetic pumping \citep{egedal2021-mixing, egedal2021-fast-limit}.

\begin{acknowledgments}
We are very grateful for discussions with Luca Comisso, Daniel Gro\v{s}elj,
Kris Klein, and Navin Sridhar.
We thank the anonymous referee for thoughtful comments and suggestions.
AT and LS were partly supported by NASA ATP 80NSSC20K0565.
AT was partly supported by NASA FINESST 80NSSC21K1383.
FL and EGZ were supported by NSF PHY 2010189.
MAR thanks support from ANID Fondecyt Regular Grant No.~1191673.
Simulations and analysis used the computer clusters Habanero, Terremoto,
Ginsburg (Columbia University), and Pleiades (NASA).
Computing resources were provided by Columbia University's Shared Research
Computing Facility (SRCF) and the NASA High-End Computing Program through the
NASA Advanced Supercomputing (NAS) Division at Ames Research Center.
Columbia University's SRCF is supported by NIH Research Facility Improvement
Grant 1G20RR030893-01 and the New York State Empire State Development, Division
of Science Technology and Innovation (NYSTAR) Contract C090171.
\end{acknowledgments}

\facility{Pleiades}

\appendix

\section{Drift-kinetic moment equations} \label{app:dk}

Here we
derive Eq.~\eqref{eq:driftkinetic} from a set of moment equations, similar
to the drift-kinetic models of \citet{zweibel2020} and \citet{ley2022}; a
more general form is given by \citet[Eqs.~31--32]{chew1956}.
Assuming gyrotropy, compression perpendicular to $\vec{B}$, and
Lorentz pitch-angle scattering
with rate $\nu$ constant over momentum and pitch angle, the relativistic Vlasov
equation is
\begin{equation} \label{eq:vlasov}
    \frac{\ptl f}{\ptl t}
    + \frac{\dot{B}}{B} \frac{p_\perp}{2} \frac{\ptl f}{\ptl p_\perp}
    = \frac{\ptl}{\ptl \mu} \left(
        \frac{\nu(1-\mu^2)}{2} \frac{\ptl f}{\ptl \mu}
    \right)
\end{equation}
where $v$ is normalized to $c$, $p$ is normalized to $m c$, and
$m$ is either ion or electron mass, depending on the species of interest.
Let us compute evolution equations for the moments
$P_\perp = \langle p_\perp v_\perp/2\rangle$
and $P_\prll = \langle p_\prll v_\prll \rangle$,
where $\langle\chi\rangle = \int \chi f \dtl^3\vec{p}$, by multiplying
Eq.~\eqref{eq:vlasov} by $p_\perp v_\perp/2$ and $p_\prll v_\prll$.
For $P_\perp$, we have:
\begin{align*}
    \frac{\dtl P_\perp}{\dtl t}
    &= - \frac{\dot{B}}{B} \int \frac{1}{4} p_\perp^2 v_\perp
        \frac{\ptl f}{\ptl p_\perp}
        2\pi p_\perp \dtl p_\perp \dtl p_\prll
        - \nu\left(P_\perp-P_\prll\right) \\
    &= \frac{\dot{B}}{B} \left\langle
        \frac{1}{2} p_\perp v_\perp \left( 2 - \frac{1}{2} v_\perp^2 \right)
        \right\rangle
        - \nu\left(P_\perp-P_\prll\right)
\end{align*}
Similarly for $P_\prll$, we have:
\begin{align*}
    \frac{\dtl P_\prll}{\dtl t}
    &= - \frac{\dot{B}}{B} \int p_\prll v_\prll
        \frac{p_\perp}{2} \frac{\ptl f}{\ptl p_\perp}
        2\pi p_\perp \dtl p_\perp \dtl p_\prll
        + 2\nu\left(P_\perp-P_\prll\right) \\
    &= \frac{\dot{B}}{B} \left\langle
            p_\prll v_\prll \left( 1 - \frac{1}{2} {v_\perp}^2 \right)
        \right\rangle
        + 2\nu\left(P_\perp-P_\prll\right)
\end{align*}
In the non-relativistic limit,
\begin{align*}
    \frac{\dtl P_\perp}{\dtl t}
        &= \frac{\dot{B}}{B} 2 P_\perp
        - \nu\left(P_\perp-P_\prll\right) \\
    \frac{\dtl P_\prll}{\dtl t}
        &= \frac{\dot{B}}{B} P_\prll
        + 2\nu\left(P_\perp-P_\prll\right)
\end{align*}
which we then use to obtain Eq.~\ref{eq:driftkinetic}.

\section{Whistler-mode offset from bi-Maxwellian dispersion} \label{app:offset}

\begin{figure*}
    \includegraphics[width=7.1in]{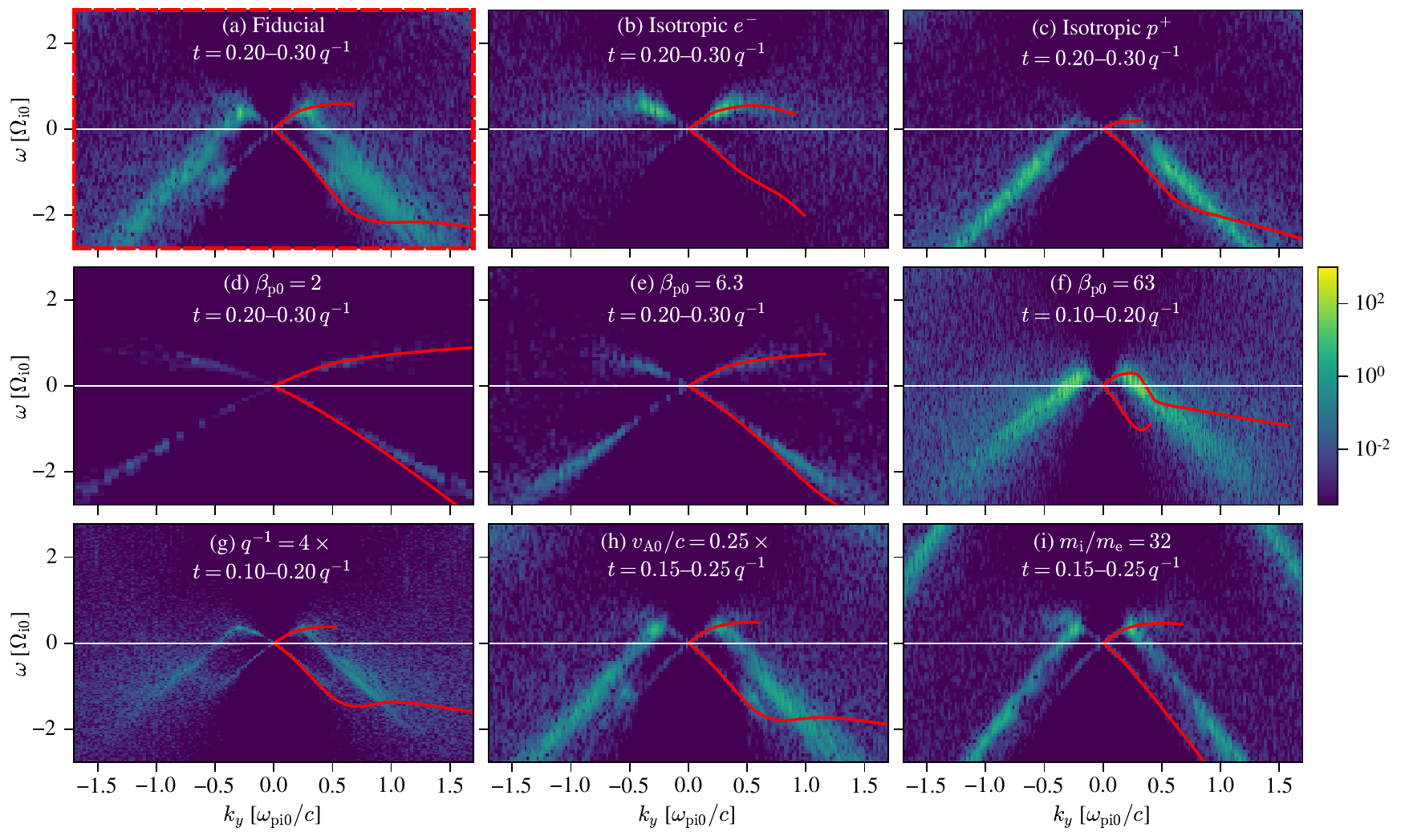}
    \caption{
        The RCP wave offset is a high-$\bp$ effect driven by anisotropic
        electrons in the presence of hot ions, which we show by plotting
        $(\omega,k)$ power spectra, at early times in the simulation when RCP
        waves are first excited, for eight simulations altered in one or a few
        parameters with respect to our fiducial simulation.
        Like Fig.~\ref{fig:fft}, RCP/LCP wave power has $\omega > 0$ and $< 0$
        respectively; red curves are whistler and IC dispersion curves,
        truncated at strong damping $\Gamma(k) < -|\omega(k)|$.
        (a) Fiducial simulation, same as Fig.~\ref{fig:fft}(a).
        (b) Isotropic electrons using $q_\mt{iso}=2q/3$.
        (c) Isotropic protons using $q_\mt{iso}=2q/3$.
        (d-f) Initial total plasma beta $\bpo=2$, $6.3$, and $63$.
        (g) Slower compression with $q^{-1}= 4\times$ larger than fiducial
        simulation.
        (h) Lower $\vAo/c$ by a factor $0.25\times$ with respect to fiducial
        simulation; i.e., less relativistic.
        (i) Raise $\mime=32$ and lower $\vAo/c$, same simulation as shown in
        Fig.~\ref{fig:energygain-vAc-mime}(h).
        Panels (f-i) have different time selections because altering $\bpo$,
        $q^{-1}$, $\vAo/c$, and $\mime$ also alters when the RCP offset waves
        appear.
    }
    \label{fig:fft-offset}
\end{figure*}

What causes the RCP mode offset discussed in Sec.~\ref{sec:wave-id}?
Though we do not yet know, we checked how it behaves in varying plasma
conditions.
The offset mode must come from free energy in electron temperature anisotropy
flowing into the whistler branch, and the offset requires hot ions, based on
several simulations shown in Fig.~\ref{fig:fft-offset}.
If we compress electrons isotropically, the offset mode disappears, whereas
if we compress ions isotropically, the offset mode persists
(Fig.~\ref{fig:fft-offset}(b-c)).
The offset persists at higher $\bpo=63$ and disappears at lower $\bpo=6.3$ and
$\bpo=2$, with $T_\mt{i0} = T_\mt{e0}$ for all $\bpo$ values
(Fig.~\ref{fig:fft-offset}(d-f)).
The offset mode persists at larger $q^{-1}$ and lower $\vAo/c$, i.e., towards
more realistic ICM conditions (Fig.~\ref{fig:fft-offset}(g-h)).
And, the offset mode persists at $\mime=32$; the location and the bandwidth of
the mode power in $(\omega,k)$ space follows the whistler branch rather than
the IC branch (Fig.~\ref{fig:fft-offset}(i)).

\section{Why do waves form two frequency bands?} \label{app:halt-compress}

We perform four numerical experiments to check the origin of the two
distinct frequency bands of wave power in Fig.~\ref{fig:overview}(a).

First, we halt the compression at $t=0.5q^{-1}$ and $t=1.0q^{-1}$.
The scale factors $a_x(t)$ and $a_z(t)$ (Eq.~\eqref{eq:scale})
are pinned to constants; the waves and particles are allowed to evolve
self-consistently without further external driving.

\begin{figure*}
    \includegraphics[width=7.1in]{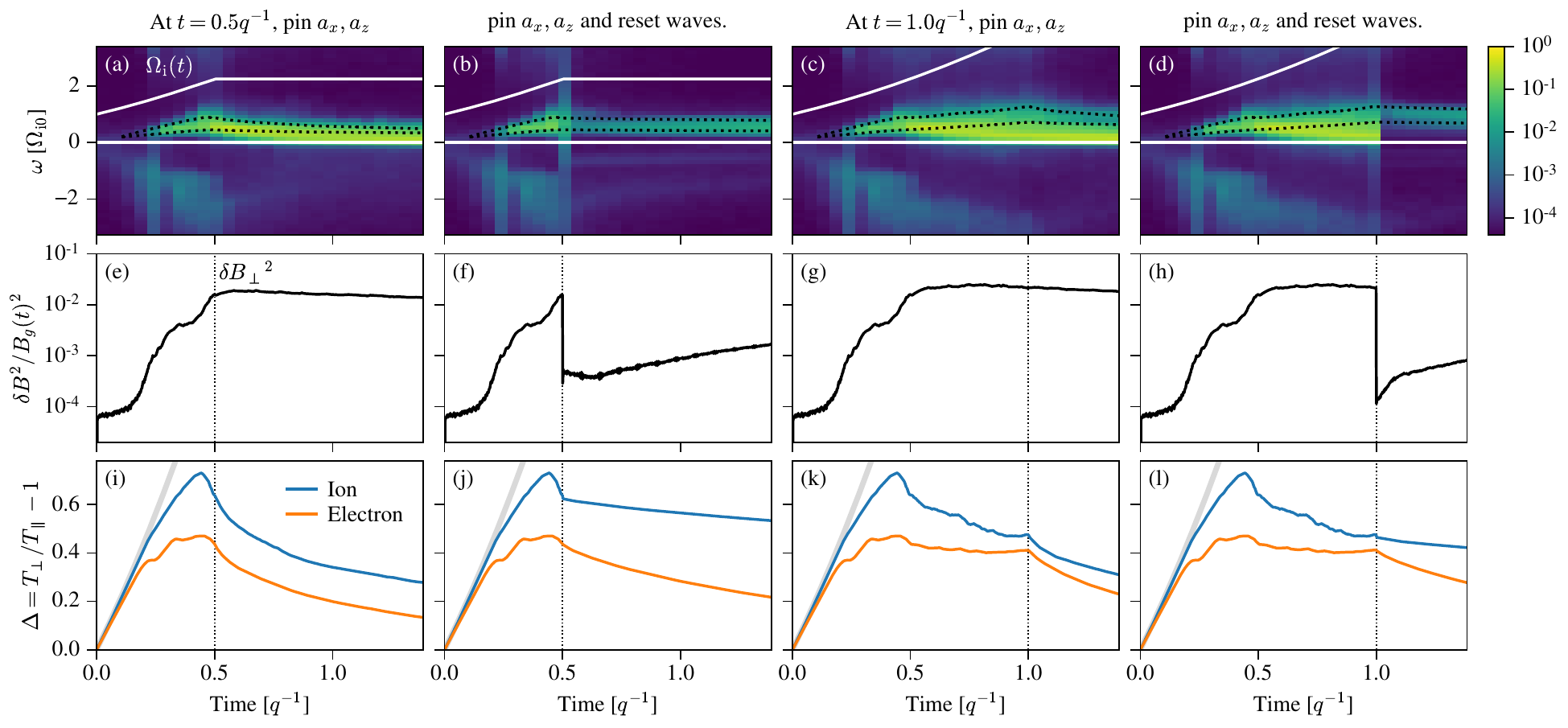}
    \caption{
        Halt compression and/or reset waves in the fiducial simulation
        (procedure given in Appendix~\ref{app:halt-compress}), to test the
        origin of low- and high-frequency LCP wave power in
        Fig.~\ref{fig:overview}(a).
        Panel layout matches Fig.~\ref{fig:overview}(a,c,d): top row is wave
        spectrogram, middle row is magnetic fluctuation power scaled to
        background field $B_g(t)$, and bottom row is ion and electron
        anisotropy $\Delta$.
        In the left two columns, compression halts at $t=0.5q^{-1}$;
        in the right two columns, compression halts at $t=1.0q^{-1}$.
        In panels (a) and (c), when compression halts, low-frequency wave power
        persists and high-frequency power weakens or does not appear.
        In panels (b) and (d), when compression halts and waves are also reset,
        the particle thermal anisotropy drives waves in the unstable frequency
        band predicted by Eq.~\eqref{eq:dispersion} (black dotted lines), and
        lower-frequency wave power does not appear.
    }
    \label{fig:halt-compress}
\end{figure*}

The result is shown by Fig.~\ref{fig:halt-compress} panels (a,c,i) and
(c,g,k).
The existing wave power drifts towards lower frequency, while the
high-frequency band either does not appear as a distinct feature
(Fig.~\ref{fig:halt-compress}(a))
or weakens in strength (Fig.~\ref{fig:halt-compress}(c)) as compared to
Fig.~\ref{fig:overview}(a).

Then, we halt compression and also ``reset'' waves to see (i) what waves
are driven unstable by particles' own anisotropic distribution, and (ii) if
said waves are reasonably predicted by the non-relativistic bi-Maxwellian
approximation of Eq.~\eqref{eq:dispersion}.
To ``reset'' waves, we zero all electromagnetic fields except for the
background field $B_g$.
We also subtract all particles' bulk motion as follows.
We compute the ion and electron bulk 3-velocities with a 5-cell kernel for
particle-to-grid mapping.
All macroparticles are Lorentz boosted so as to cancel their own species'
bulk velocity; their PIC weights are also adjusted to account for the
spatial part of the Lorentz transformation \citep{zenitani2015}.
The velocity subtraction is not perfect; it leaves a residual bulk motion
at a few percent of its original amplitude.
So, we apply the same velocity subtraction procedure again.
Two velocity subtractions suffice to leave no detectable ion bulk motion.

The result of halting compression and resetting waves is shown by
Fig.~\ref{fig:halt-compress} panels (b,f,j) and (d,h,l).
The anisotropic particle distributions grow waves in a comparatively
``high'' frequency band consistent with the unstable wave prediction of
Eq.~\eqref{eq:dispersion}.

\section{Scattering measurement timestep} \label{app:idt}

To measure pitch-angle scattering in Fig.~\ref{fig:scattering}, the
measurement timestep $\Delta t$ cannot be too short or too long.

If $\Delta t$ is too short, an electron may not have time to interact with
one or multiple waves; its trajectory in momentum space may not yet be
diffusive.
The relativistic cyclotron frequency $e B/(\gamma\me c) \sim \Omega_\mt{i}$
for $p/(\me c) \sim 10$ and $\mi/\me = 8$, so a timestep $\Delta t \gtrsim$
a few ${\Omcio}^{-1}$ should suffice to resolve the wave-particle
interaction.  More energetic electrons with larger $\gamma$ and hence
slower gyration may need a correspondingly longer timestep.

If $\Delta t$ is too long, electrons may scatter out of the wave resonance
and experience very different scattering rates within the measurement time
$\Delta t$; our measurement becomes non-local in $\mu$.
The wave resonance region itself may evolve in time.
And, electron displacements in $\mu$ may become comparable to the finite
range of $\mu \in [-1,1]$; our measurement of
$\langle\Delta\mu\Delta\mu\rangle$ would trend towards a constant rather
than increasing linearly with $\Delta t$ as expected for an unbounded
random walk.

\begin{figure*}
    \plotone{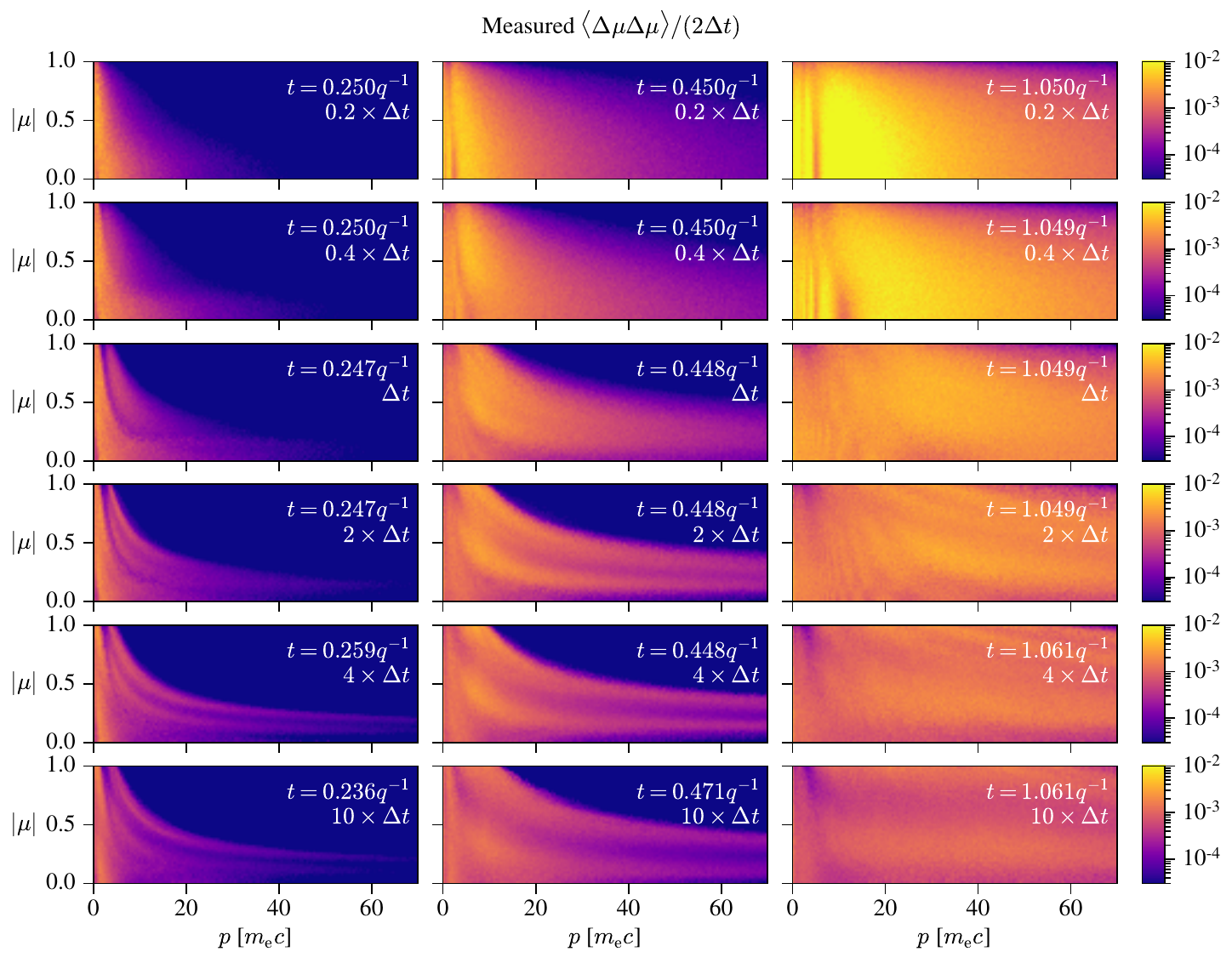}
    \caption{
        Effect of varying $\Delta t$ upon the measured pitch-angle
        scattering rate, measured between times $t$ and $t+\Delta t$.
        The scattering timestep $\Delta t$ is smallest at top and increases
        going down each row, varying from $0.9$ to $47 {\Omcio}^{-1}$;
        the third row from the top corresponds to
        $\Delta t = 4.7{\Omcio}^{-1}$ as used for
        Fig.~\ref{fig:scattering}(j,k,l).
        The simulation time varies from left to right columns as
        $t\approx 0.25q^{-1}$, $0.45q^{-1}$ and $1.05q^{-1}$ to match
        Fig.~\ref{fig:scattering}.
    }
    \label{fig:idt-vary}
\end{figure*}

In Fig.~\ref{fig:idt-vary}, we show how altering $\Delta t$ by $0.2\times$
to $10\times$ (i.e., $0.9{\Omcio}^{-1}$ to $47{\Omcio}^{-1}$) then alters
the measured scattering rates
$\langle\Delta\mu\Delta\mu\rangle/(2\Delta t)$ in phase space coordinates
$(p,\mu)$.
Recall that our fiducial $\Delta t = 4.7{\Omcio}^{-1}$ in
Fig.~\ref{fig:scattering}.

\section{Numerical convergence} \label{app:conv}

\begin{figure*}
    \centering
    \includegraphics[scale=1]{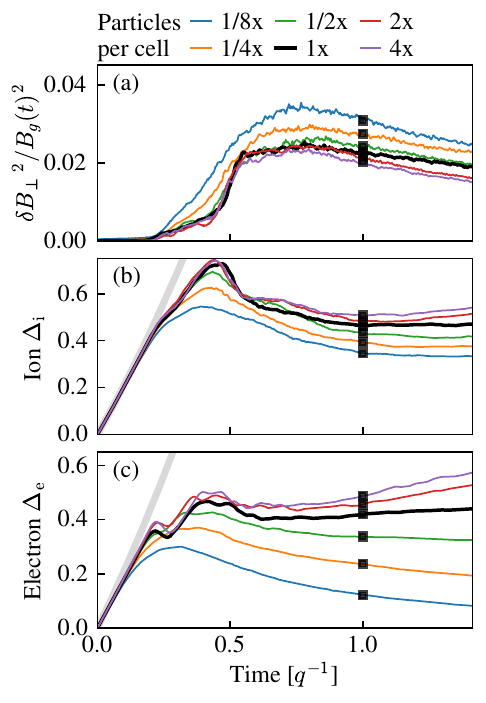}
    \includegraphics[scale=1]{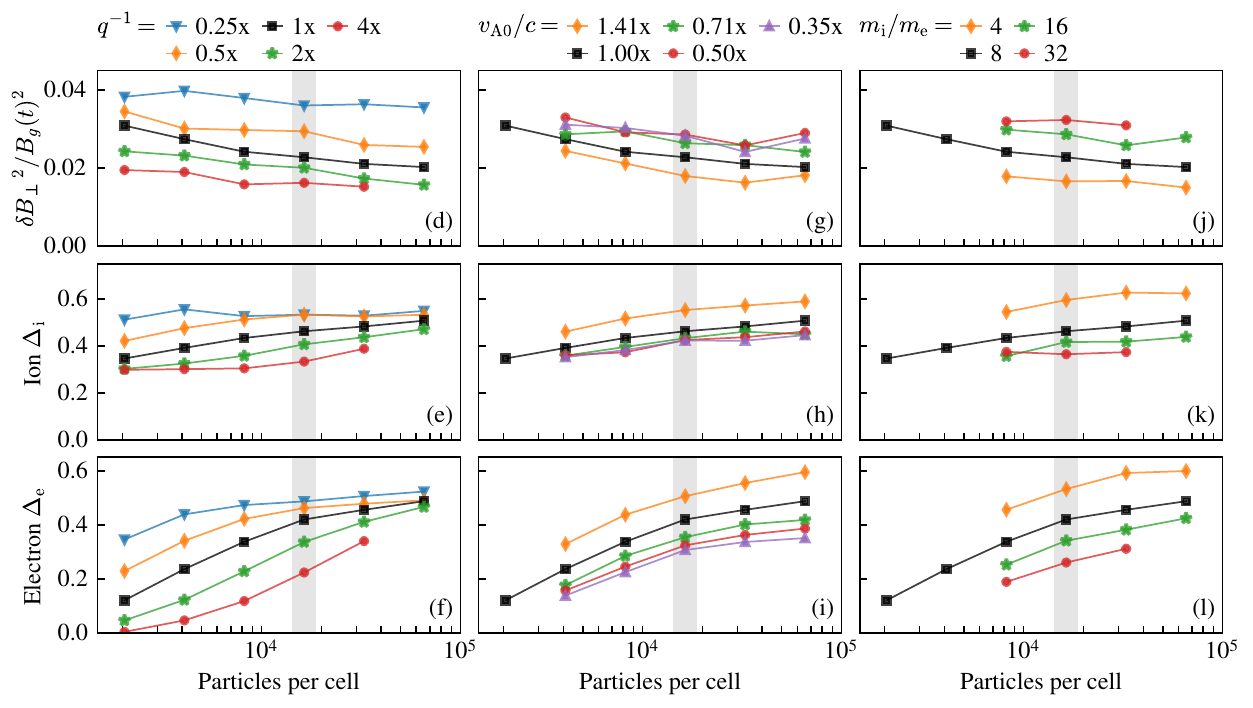}
    \caption{
        Convergence of our simulations with respect to particle sampling for
        thermal ICM particles (both ions and electrons).
        Left-most column: time evolving (a) wave power $\delta {B_\perp}^2$,
        (b) ion anisotropy $\Delta_\mt{i}$, and (c) electron anisotropy
        $\Delta_\mt{e}$ for varying particles per cell (colored curves)
        compared to our fiducial simulation (black curve).
        Thick light-gray curve is non-relativistic CGL prediction in (b-c).
        Black squares at $t=1 q^{-1}$ correspond to same symbols in (d-f).
        Right three columns: wave power and anisotropy, measured at $t=1
        q^{-1}$ for simulations with varying $q^{-1}$ (d-f), $\vAo/c$ (g-i),
        and $\mime$ (j-l).
        Each marker set represents one simulation from the main manuscript with
        varied particle sampling.
        Black squares represent fiducial simulation in all panels (d-l), and
        correspond to the data and markers in (a-c).
        Vertical light-gray bar indicates fiducial particle sampling of
        $16,384$ ions and electrons per cell (excluding test-particle CRe);
        all markers within light-gray bar correspond to a simulation from the
        main manuscript (cf.~Table~\ref{tab:param}).
        Legends above each column report ratio of $q^{-1}$, $\vAo/c$, and
        $\mime$ with respect to fiducial simulation ($1\times$).
    }
    \label{fig:ppc}
\end{figure*}

In Fig.~\ref{fig:ppc} we show numerical convergence with respect to particles
per cell, focusing on total wave power $\delta {B_\perp}^2$, ion temperature
anisotropy $\Delta_\mt{i}$, and electron temperature anisotropy $\Delta_\mt{e}$.
In particular, we sample these quantities at $t=1 q^{-1}$ in order to check
convergence at late times when waves scatter CRe appreciably.
We check convergence for our fiducial simulation, and also all runs with
varying $q^{-1}$, $\vAo/c$, and $\mime$.
The simulations in Fig.~\ref{fig:ppc} used single-precision floats for
particle momenta in the PIC algorithm, which introduces a small numerical
error (see Sec.~\ref{sec:methods}).
This precision error does not depend on particle sampling, so we consider
it acceptable for our convergence test.

It's most important that the wave power and ion temperature anisotropy are
converged with respect to particle sampling for our study.
For all simulations considered, a two or four times increase in particle count
does not modify $\delta {B_\perp}^2$ or $\Delta_\mt{i}$ by more than a factor
of $1.5\times$.
We consider this rate of convergence acceptable.

The electron temperature anisotropy is more sensitive to particle sampling.
Some simulations are not converged in $\Delta_\mt{e}$, particularly those with
large $q^{-1}$.
We consider this incomplete convergence acceptable because of the minor role of
electron-driven waves in CRe energization, as shown by our simulations of CRe
energy gain with electrons heated isotropically to prevent whistler wave growth
(Fig.~\ref{fig:energygain-vAc-mime}(i-l)).

\section{Simulation Parameters} \label{app:param}

Table~\ref{tab:param} provides input parameters for all simulations in this
manuscript: first the fiducial simulation, followed by parameter sweeps of
$q^{-1}$, $\kB T_0$ (equivalently $\vAo/c$), $\mime$, and $\beta_\mt{p0}$.
The simulations with varying $\beta_\mt{p0}$ are only used in
Appendix~\ref{app:offset}.
Simulations with varying particle count (Appendix~\ref{app:conv}) or with one
species isotropic are not explicitly shown.

We define some input parameters in code units:
\texttt{my} is the domain size in cells;
\texttt{intv} is the number of timesteps between output file dumps, relevant
for wave power spectra and particle scattering measurements;
\texttt{dur} is the simulation duration in timesteps.
Other key parameters such as grid cell size, particles per cell, current
filtering, and numerical speed of light are identical across all simulations
and are stated in Sec.~\ref{sec:methods}.

\begin{deluxetable*}{lrrrrrrrrrrr}
\tabletypesize{\scriptsize}
\tablecaption{
    Simulation input parameters.
    Columns are defined in Sec.~\ref{sec:methods} and Appendix~\ref{app:param}.
    \label{tab:param}
}
\tablehead{
    \multicolumn{1}{l}{Purpose}  
      & \multicolumn{1}{r}{$\mi/\me$}  
      & \multicolumn{1}{r}{$\beta_\mt{p0}$}
      & \multicolumn{1}{r}{$\kB T_0$}
      & \multicolumn{1}{r}{$\vAo/c$}
      & \multicolumn{1}{r}{$q^{-1}$}
      & \multicolumn{1}{r}{\texttt{my}}
      & \multicolumn{1}{r}{\texttt{my}}
      & \multicolumn{1}{r}{\texttt{intv}}
      & \multicolumn{1}{r}{\texttt{intv}}
      & \multicolumn{1}{r}{\texttt{dur}}
      & \multicolumn{1}{r}{\texttt{dur}} \\
    \multicolumn{1}{l}{}  
      & \multicolumn{1}{r}{}
      & \multicolumn{1}{r}{}
      & \multicolumn{1}{r}{$\left[\me c^2\right]$}
      & \multicolumn{1}{r}{}
      & \multicolumn{1}{r}{$\left[{\Omcio}^{-1}\right]$}
      & \multicolumn{1}{r}{}
      & \multicolumn{1}{r}{$\left[\rLio\right]$}
      & \multicolumn{1}{r}{}
      & \multicolumn{1}{r}{$\left[{\Omcio}^{-1}\right]$}
      & \multicolumn{1}{r}{}
      & \multicolumn{1}{r}{$\left[q^{-1}\right]$}
}
\startdata
Fiducial & 8 & 20.0 & 0.20 & 0.067 & 800 & 4608 & 79.3 & 800 & 0.94 & 960000 & 1.41 \\
\hline
Vary $q^{-1}$ & 8 & 20.0 & 0.20 & 0.067 & 200 & 4608 & 79.3 & 800 & 0.94 & 240000 & 1.41 \\
Vary $q^{-1}$ & 8 & 20.0 & 0.20 & 0.067 & 400 & 4608 & 79.3 & 800 & 0.94 & 480000 & 1.41 \\
Vary $q^{-1}$ & 8 & 20.0 & 0.20 & 0.067 & 1600 & 4608 & 79.3 & 800 & 0.94 & 1920000 & 1.41 \\
Vary $q^{-1}$ & 8 & 20.0 & 0.20 & 0.067 & 3200 & 4608 & 79.3 & 800 & 0.94 & 3840000 & 1.41 \\
\hline
Vary $v_\mathrm{A0}/c$ & 8 & 20.0 & 0.40 & 0.094 & 800 & 4608 & 79.3 & 600 & 1.00 & 720000 & 1.50 \\
Vary $v_\mathrm{A0}/c$ & 8 & 20.0 & 0.10 & 0.047 & 800 & 4608 & 79.3 & 1200 & 1.00 & 1440000 & 1.50 \\
Vary $v_\mathrm{A0}/c$ & 8 & 20.0 & 0.05 & 0.033 & 800 & 4608 & 79.3 & 1700 & 1.00 & 2040000 & 1.50 \\
Vary $v_\mathrm{A0}/c$ & 8 & 20.0 & 0.03 & 0.024 & 800 & 4608 & 79.3 & 2400 & 1.00 & 2880000 & 1.50 \\
\hline
Vary $m_\mathrm{i}/m_\mathrm{e}$ & 4 & 20.0 & 0.20 & 0.089 & 800 & 3840 & 88.7 & 400 & 0.89 & 480000 & 1.34 \\
Vary $m_\mathrm{i}/m_\mathrm{e}$ & 16 & 20.0 & 0.20 & 0.049 & 800 & 6144 & 77.0 & 1600 & 0.97 & 1920000 & 1.46 \\
Vary $m_\mathrm{i}/m_\mathrm{e}$ & 32 & 20.0 & 0.20 & 0.035 & 800 & 9216 & 82.8 & 3200 & 0.98 & 3840000 & 1.48 \\
\hline
Vary $\beta_\mathrm{p0}$ & 8 & 2.0 & 0.20 & 0.211 & 800 & 1536 & 83.6 & 300 & 1.12 & 360000 & 1.68 \\
Vary $\beta_\mathrm{p0}$ & 8 & 6.3 & 0.20 & 0.119 & 800 & 2688 & 82.3 & 500 & 1.05 & 600000 & 1.57 \\
Vary $\beta_\mathrm{p0}$ & 8 & 63.2 & 0.20 & 0.037 & 800 & 8192 & 79.3 & 1500 & 0.99 & 1800000 & 1.49
\enddata  
\tablecomments{
Table~\ref{tab:param} is
available in a machine-readable CSV format in the online journal.
}
\end{deluxetable*}

\bibliographystyle{aasjournal}
\bibliography{library}

\end{document}